\documentclass[11pt,a4paper]{article}
\usepackage{amsmath}
\usepackage{amsthm}
\usepackage{amssymb}
\usepackage{graphicx}

\begin{document}

\newcommand{\beq}[1]{\begin{equation}\label{#1}}
\newcommand{\eq}{\end{equation}}

\newtheorem{predl}{Proposition}%[section]
\newtheorem{lem}{Lemma}[section]
\newtheorem{theor}{Theorem}[section]

\begin{flushright}
ITEP-TH-43/13
\end{flushright}

\begin{center}

\vspace{20mm}

{\Large{\bf Semi-invariants and Integrals\\ of the Full Symmetric $\mathfrak{sl}_n$ Toda Lattice}
}\\

\vspace{5mm}

{\sc Yu.B. Chernyakov${}^{\; a}$ and A.S. Sorin$^{\; b}$}\\[15pt]

{${}^a$\small Institute for Theoretical and Experimental Physics\\ Bolshaya Cheremushkinskaya, 25, 117218 Moscow, RUSSIA and\\ Bogoliubov Laboratory of Theoretical Physics\\ Joint Institute for Nuclear Research\\ 141980 Dubna, Moscow region,  RUSSIA.\\}e-mail: {\small \it
chernyakov@itep.ru}\vspace{10pt}

{${}^b$\small Bogoliubov Laboratory of Theoretical Physics and\\
Veksler and Baldin Laboratory of High Energy Physics\\
Joint Institute for Nuclear Research \\ 141980 Dubna, Moscow Region, RUSSIA
\\ }e-mail: {\small \it
sorin@theor.jinr.ru}\vspace{10pt}
\vspace{3mm}

\end{center}

\begin{abstract}
We consider the full symmetric version of the Lax operator of the Toda lattice which is known as the full symmetric Toda lattice.
The phase space of this system is the generic orbit of the coadjoint action of the Borel subgroup $B^+_n$ of $SL_n(\mathbb R)$.
This system is integrable. We propose a new method of constructing semi-invariants and integrals of the full symmetric Toda lattice. Using only the Toda equations for the Lax eigenvector matrix we prove the existence of the semi-invariants which are Pl\"ucker coordinates in the corresponding projective spaces. Then we use these semi-invariants to construct the integrals. It is known that the full symmetric $\mathfrak{sl}_n$ Toda lattice has additional integrals which can be produced by Kostant procedure except for the integrals which can be derived by the chopping procedure. Altogether these integrals constitute a full set of the independent non-involutive integrals. Yet the unsolved complicated technical problem is their explicit derivation since Kostant procedure has crucial computational complexities even for low-rank Lax matrices and is practically unapplicable for higher ranks. Our new approach provides a resolution of this problem and results in simple explicit formulae for the full set of independent semi-invariants and integrals expressed both in terms of the Lax matrix and its eigenvector and eigenvalue matrices of the full symmetric $\mathfrak{sl}_n$ Toda lattice without using the chopping and Kostant procedures. We also describe the structure of the additional integrals of motion as functions on the flag space modulo the Toda flows and show how Pl\"ucker coordinates of different projective spaces define different families of the additional integrals. In this paper we present detailed proofs of the propositions of \cite{CS}.

\end{abstract}

\section{Introduction}
Non-periodic Toda lattice (Toda chain) consists of $n$ particles on the line with interactions between neighbours.
The Hamiltonian of this system is given by
\beq{H-Toda}
H = \sum^{n}_{i=1}\frac{1}{2}p_{i}^{2} + \sum^{n-1}_{i=1} \exp(q_{i}-q_{i+1}),
\eq
where $p_{i}$ is the momentum of the $i^{th}$ particle and $q_{i}$ is its coordinate.
The Poisson structure on the phase space $(p_{i}, q_{i})$ has the well-known form
\beq{Poiss-str-1}
\{ p_{i}, q_{j} \} = \delta_{ij}, \ \{ p_{i}, p_{j} \} = 0, \ \{ q_{i}, q_{j} \} = 0.
\eq
The evolution of the system is given by the standard Hamiltonian equations: $p'_i=\{H,\,q_i\},\ q'_i=-\{H,\,p_i\}$.

If we make the following ansatz ( \cite{F1})
\beq{Var}
b_{i} = p_{i}, \ \ \ a_{i} = \exp \frac{1}{2}(q_{i}-q_{i+1}),
\eq
the Hamiltonian will take the form
\beq{H-Toda-2}
H = \sum^{n}_{i=1}\frac{1}{2}b_{i}^{2} + \sum^{n-1}_{i=1} a_{i}^{2}.
\eq
Observe, that there are only $n-1$ variables $a_i$. The Poisson structure (\ref{Poiss-str-1}) turns into
\beq{Poiss-str-2}
\{ b_{i}, a_{i-1} \} = -a_{i-1}, \ \{ b_{i}, a_{i} \} = a_{i}.
\eq
All the other brackets of coordinates $a_i$ and $b_j$ are equal to zero.
In these coordinates it is easy to find the Lax representation of the system:
 one can show that the Hamilton equations are equivalent to the following matrix equation.
\beq{LM}
\dot{L} = [B,L],
\eq
where $L$ is called the \textit{Lax operator} and is given by
\beq{Lax} L = \left(
\begin{array}{c c c c c c}
 b_{1} & a_{1} & 0 & ... &0\\
 a_{1} & b_{2} & a_{2} &... & 0\\
 0 & ... & ... & ...& 0\\
 0 & ... & a_{n-2} & b_{n-1} &a_{n-1}\\
 0 & 0 & ... &  a_{n-1} & b_{n}\\
\end{array}
\right)
\eq
and $B$ is the operator
\beq{Lax2} B = \left(
\begin{array}{c c c c c c}
 0 & -a_{1} & 0 & ... &0\\
 a_{1} & 0 & -a_{2} &... & 0\\
 0 & ... & ... & ...& 0\\
 0 & ... & a_{n-2} & 0 & -a_{n-1}\\
 0 & 0 & ... & a_{n-1} & 0\\
\end{array}
\right).
\eq
The equation (\ref{LM}) is the compatibility condition of the system:
\beq{compat-Lax}
\left\{
\begin{array}{c}
L \Psi = \Psi \Lambda,\\
\ \\
\frac{\partial}{\partial t} \Psi = B \Psi,
\end{array}
\right.
\eq
where $\Psi$ is the eigenvector matrix of the Lax operator and $\Lambda$ is the eigenvalue matrix of the Lax operator.

This system was first considered in \cite{T1, T2}, and in the work \cite{H} there were found $n$ functionally independent integrals of the motion. The involution of the integrals was proved in the papers \cite{F1, F2}. In particular, it is easy to see, that the trace of $L$ is a Casimir function of the system, so that we can set it equal to $0$, without any loss of generality. Thus we can assume, that $Tr(L)=0$.

It turns out, that this system is Hamiltonian: one can regard it as the dynamical system on the orbits of the coadjoint action of the Borel subgroup $B^+_n$ of $SL_n(\mathbb R)$ (equal to the group of upper triangular matrices with determinant equal to $1$) see \cite{A, Ad, K1, S}.

It is possible to realize the phase space of the dynamical system in two different ways identifying the algebra $\mathfrak{sl}_n$ with its dual using Killing form on $\mathfrak{sl}_n$ and so to generalize the Toda lattice (tri-diagonal Toda chain). The first way:
\beq{Decomp-1}
\begin{array}{c}
\mathfrak{sl}_n=\mathfrak{n}^{-}_{n} \oplus \mathfrak{b}^+_n,\\
\mathfrak{sl}^{\ast}_{n} = (\mathfrak{b}^{+}_n)^{\ast} \oplus (\mathfrak{n}^{-}_{n})^{\ast} \cong \mathfrak{b}^{-} \oplus \mathfrak{n}_n^{+},\\
(\mathfrak{b}^{+}_n)^{\ast} \cong (\mathfrak{n}^{-}_n)^{\perp} = \mathfrak{b}^-_n, \ \ \ (\mathfrak{n}^{-}_n)^{\ast} \cong (\mathfrak{b}^{+}_n)^{\perp} = \mathfrak{n}_n^{+}
\end{array}
\eq
where $\mathfrak{b}^+_n$ and $\mathfrak{b}^-_n$ are the algebras of upper triangular and lower triangular matrices accordingly, а $\mathfrak{n}_n^{+}$ and $\mathfrak{n}^{-}_n$ are the algebras of strictly upper triangular and strictly lower triangular matrices respectively. Now we can identify the phase space of the Toda lattice with the orbit of the coadjoint action of the Borel subgroup $B^+_n$ in the affine space $\mathfrak{b}^-_n + \epsilon$ (where $\epsilon$ is the sum of simple roots).

Or else we can use the following identification:
\beq{Decomp-2}
\begin{array}{c}
\mathfrak{sl}_n=\mathfrak{so}_n \oplus \mathfrak{b}^+_n,\\
\mathfrak{sl}^{\ast}_{n} = (\mathfrak{b}^{+}_n)^{\ast} \oplus (\mathfrak{so}_n)^{\ast} \cong Symm_n \oplus \mathfrak{n}_n^{+},\\
(\mathfrak{b}^{+}_n)^{\ast} \cong (\mathfrak{so}_n)^{\perp} = Symm_n, \ \ \ (\mathfrak{so}_n)^{\ast} \cong (\mathfrak{b}^{+}_n)^{\perp} = \mathfrak{n}_n^{+}
\end{array}
\eq
As one sees we can identify the space of symmetric matrices with the dual space of Lie algebra of Borel subgroup: $Symm_n\cong(\mathfrak{b}_n^+)^*$, and hence we can introduce a symplectic structure on $Symm_n$, pulling it back from $(\mathfrak{b}_n^+)^*$.

Based on these two approaches we get two possible generalizations of the Toda chain which lead to two integrable systems called the Full Kostant-Toda lattice and the Full Symmetric Toda lattice.

The dimensions of the phase space of these systems are greater than $2n$ so for the integrability we need more integrals of motion than in the case of the tri-diagonal Toda lattice. First the additional integrals for the Full Symmetric Toda lattice were found in paper \cite{DLNT} where the  so called chopping procedure was defined. The constructions of the additional integrals and semi-invariants used in them were explored in papers \cite{DLNT, A, FS, EFS}.
In recent paper \cite{T} the formulae for the integrals without the chopping procedure were found. Further this formulae were used to generalization of the Toda lattice in the quantum case.

In paper \cite{EFS} it was shown that the integrals of motion of the Full Kostant-Toda lattice can be expressed via homogeneous coordinates of some projective spaces. These expressions come from the embedding in the flag space and dynamics with respect to 1-parametric subgroups induced by the iso-spectral integrals of motion (see Appendix B). However, in the case of the Full Symmetric Toda lattice the mapping from the space of symmetric matrices to flag space is not embedding.

So, in our paper we calculate the dynamics of the matrix elements and minors of the matrix of the eigenvectors of the Lax operator using the equations of motion.

First the existence of two different families of the integrals in involution of the Full Kostant-Toda lattice was found for the case $n=4$ in paper \cite{EFS}. In papers \cite{BG-1, GS, BG-2} it was shown that this system is integrable in non-commutative sense (see \cite{Neh} on non-commutative integrability). The full set of the integrals of motion of the Full Symmetric Toda lattice consists of iso-spectral integrals, integrals obtained by chopping procedure and additional integrals.

\paragraph{The aims of our paper}
are to calculate the dynamics of the semi-invariants of the motion, to find an explicit form of the integrals of motion and to describe their structure.

In chapter 2 we give a description of the integrability of the Full Symmetric Toda lattice and describe the chopping procedure as defined in the paper \cite{DLNT}.

In chapters 3, 4, 5 the main results are given.

In conclusion we give a short characterization of the results and give the acknowledgments.

In Appendices we give the description of the flag space and the dynamics on the flag space.

\paragraph{The short description of the results.}
\ \\
In section 3 the dynamics of the minors $M$ of the eigenvector matrix $\Psi$ of the Lax operator is calculated. These minors are obtained by the intersection of the first $k<n$ rows or the last $l<n$ rows with any set of $k$ or $l$ columns. It was shown that these minors (called "good") are the semi-invariants with respect to the flows of the Full Symmetric Toda lattice which induced by iso-spectral integrals of motion:
$$\frac{\partial}{\partial t_{d-1}}M_{\frac{1,2,...,k}{i_{1},i_{2},...,i_{k}}} = (-\sum_{j=1}^{k}a^{(d-1)}_{jj} + \sum_{i_{m}=i_{1}}^{i_{k}}\lambda^{d-1}_{i_{m}}) M_{\frac{1,2,...,k}{i_{1},i_{2},...,i_{k}}} \ \ ,$$
$$\frac{\partial}{\partial t_{d-1}}\tilde{M}_{\frac{n-l+1,...,n}{i_{1},i_{2},...,i_{l}}} = (\sum_{j=n-l}^{n}a^{(d-1)}_{jj} - \sum_{i_{m}=i_{1}}^{i_{l}}\lambda^{d-1}_{i_{m}}) \tilde{M}_{\frac{n-l+1,...,n}{i_{1},i_{2},...,i_{l}}},$$
where $a^{(d-1)}_{jj}$ is matrix elements of the matrix $L^{d-1}=\underbrace{L \cdot L \cdot ... \cdot L}_{d-1}$. Here vector field $\frac{\partial}{\partial t_{d-1}}$ is induced by the iso-spectral invariant of motion $\frac{1}{d}TrL^{d}, \ d=\overline{2,n} \ $.\\
\ \\
Further, in section 4 and 5 a method to calculate the integrals of motion is suggested. This method is based on the existing of the semi-invariants of motion. The integrals of motion are rational functions of the following form:
$$
J_{k_{1},k_{2}}= \frac{A^{(k_{1})}_{\frac{n-m+1,...,n}{1,2,...,m}}}{A^{(k_{2})}_{\frac{n-m+1,...,n}{1,2,...,m}}},
$$
where $A^{(k_{1})}, \ A^{(k_{2})}$ are the semi-invariants and minors of the matrices $L^{k_{1}}, \ L^{k_{2}}$ respectively, $k_{1} \neq k_{2}, \ k_{1}, \ k_{2} \in \mathbb{N}$. This formulae allow one to calculate explicitly all additional integrals of motion.\\
\ \\
In section 5 the structure of all additional integrals of motion is described. It is shown that these integrals are rata of two semi-invariants of motion. Each of these semi-invariants consists of the sum whose terms is formed by the product of two "good" \ minors $M$ of the matrix $\Psi$ with the coefficients which are the polynomials of eigenvalues $\lambda_{i}$ of the Lax operator. Also, the structure of the full non-commutative set of the integrals of motion of the Full Symmetric Toda lattice is described. We show that the number of the additional integrals (not in the set of the integrals obtained by chopping procedure and iso-spectral integrals) equal to the number of integrals obtained by chopping procedure. In the cases $n=4$ and $n=5$ two families in involution are described in an explicit form as the functions of matrix elements of the Lax operator .\\
\ \\
Finally we give the formula which determines the size of a full non-commutative set of the integrals of motion of the Full Symmetric Toda lattice. It is the same one as in the papers \cite{BG-1, GS, BG-2}, but our approach to this formula is based on the consideration of the flag space which is described as Pl\"ucker's embedding so that the minors $M$ are the homogeneous coordinates in the corresponding projective spaces.

\section{Chopping procedure in the Full Symmetric Toda lattice}
In this section we follow the paper \cite{DLNT}. The matrix of the Lax operator of the Full Symmetric Toda lattice, real symmetric matrix of order $n$, has the following form
\beq{Lax} L = \left(
\begin{array}{c c c c c c}
 a_{11} & a_{12} & ... & a_{1n}\\
 a_{12} & a_{22} & ... & a_{2n}\\
 ... & ... & ... & ...\\
 a_{1n} & a_{2n} & ... & a_{nn}\\
\end{array}
\right).
\eq

One can define the following set of characteristic polynomials:
\beq{Pkn}
\begin{array}{c}
P_{k}(L,\mu) = det(L - \mu I)_{k},\\
\ \\
P_{k}(L,\mu) = \sum_{m=0}^{n-2k} E_{m,k}(L)\mu^{n-2k-m}, \,\,\, 0 \leq k \leq [n/2],
\end{array}
\eq
where $(L - \mu I)_{k}$ is the matrix of order $n-k$, which we obtain by chopping $k$ upper rows and $k$ right column of matrix $(L - \mu I)$, $[ \ \ \ ]$ means the integer part. This procedure was called \textbf{chopping procedure} in paper \cite{DLNT}. It was shown there that the functions
\beq{Integrals}
I_{m,k} = \frac{E_{m,k}(L)}{E_{0,k}(L)}, \ \ \ 0 \leq k \leq [\frac{1}{2}(n-1)],  \ \ \ 1 \leq m \leq n-2k
\eq
define $[\frac{1}{4} n^{2}]$ integrals in involution for the Full Symmetric Toda lattice on the generic orbit of order $2[\frac{1}{4} n^{2}]$. These integrals are functionally independent.

\section{Dynamics of minors of the eigenvector matrix $\Psi$}

In this section we calculate in an explicit form the dynamics of the matrix elements and minors of the eigenvector matrix $\Psi$ of the Lax operator. The results will be used to define the semi-invariants of motion from which the integrals of motion will be constructed.\\

As one notes in the introduction the motivation to calculate in an explicit form the dynamics of the matrix elements and minors of the eigenvector matrix is the fact that the mapping from the space of the symmetric matrices to the flag space is not the embedding (see Appendix B). So, we cannot calculate the dynamics on the flag space as in the case of the Full Kostant-Toda lattice.

Let us consider the action of the one-parametric subgroup induced by the iso-spectral integral of motion $\frac{1}{2}TrL^{2}.$

The equation of motion for the ($n \times n$) eigenvector matrix which generalizes the second equation in  (\ref{compat-Lax}) has the following form:

\beq{psi}
\Psi^{'} = B \Psi,
\eq
where $B$:

\beq{Bn4}
B= L_{>0} - L_{<0} = (\psi \Lambda \psi^{T})_{>0} - (\psi \Lambda \psi^{T})_{<0} =
\left(
\begin{array}{c c c c c c}
 0 & a_{12} & a_{13} & ... & ... & a_{1n}\\
 -a_{12} & 0 & a_{23} & ... & ... & a_{2n}\\
 -a_{13} & -a_{23} & 0 & ... & ... & a_{3n}\\
 ... & ... & ... & ... & ... & ...\\
 ... & ... & ... & ... & ... & a_{n-1 n}\\
 -a_{1n} & -a_{2n} & -a_{3n} & ... & -a_{n-1 n} & 0\\
\end{array}
\right) \eq

After transformations we get:
$$ \left(
\begin{array}{c c c c c c}
 \psi_{11} & \psi_{12} & ... & \psi_{1n}\\
 \psi_{21} & \psi_{22} & ... & \psi_{2n}\\
 ... & ... & ... & ...\\
 \psi_{n1} & \psi_{n2} & ... & \psi_{nn}\\
\end{array}
\right)^{'} = $$

\beq{N-2} = \left(
\begin{array}{c c c c}
   (-a_{11} + \lambda_{1})\psi_{11} & (-a_{11} + \lambda_{2})\psi_{12} & ... & (-a_{11} + \lambda_{n})\psi_{1n}\\
       &  &  & \\
(-a_{22} + \lambda_{1})\psi_{21} - & (-a_{22} + \lambda_{2})\psi_{22} - & ... & (-a_{22} + \lambda_{n})\psi_{2n} -\\
       - 2a_{12}\psi_{11} & - 2a_{12}\psi_{12} &  & - 2a_{12}\psi_{1n}\\
       &  &  & \\
... & ... & ... & ...\\
       &  &  & \\
(-a_{n-1n-1} + \lambda_{1})\psi_{n-11} - & (-a_{n-1n-1} + \lambda_{2})\psi_{n-12} - & ... & (-a_{n-1n-1} + \lambda_{n})\psi_{n-1n} -\\
- 2(a_{1n-1}\psi_{11} +... & - 2(a_{1n-1}\psi_{12} +... & ... & - 2(a_{1n-1}\psi_{1n} +...\\
...+ a_{n-2n-1}\psi_{n-21}) & ...+ a_{n-2n-1}\psi_{n-22}) & ... & ...+ 2a_{n-2n-1}\psi_{n-2n})\\
       &  &  & \\
   (a_{nn} - \lambda_{1})\psi_{n1} & (a_{nn} - \lambda_{2})\psi_{n2} & ... & (a_{nn} - \lambda_{n})\psi_{nn}\\
\end{array}
\right) = \eq

$$ = \left(
\begin{array}{c c c c}
   (-a_{11} + \lambda_{1})\psi_{11} & (-a_{11} + \lambda_{2})\psi_{12} & ... & (-a_{11} + \lambda_{n})\psi_{1n}\\
       &  &  & \\
(a_{22} - \lambda_{1})\psi_{21} + & (a_{22} - \lambda_{2})\psi_{22} + & ... & (a_{22} - \lambda_{4})\psi_{2n} +\\
 + 2(a_{23}\psi_{31} +... &  + 2(a_{23}\psi_{32} +... & ... & + 2(a_{23}\psi_{3n} +...\\
...+ a_{2n}\psi_{n1}) &  ...+ a_{2n}\psi_{n2}) & ... & ...+ a_{2n}\psi_{nn})\\
       &  &  & \\
... & ... & ... & ...\\
       &  &  & \\
(a_{n-1n-1} - \lambda_{1})\psi_{n-11} + & (a_{n-1n-1} - \lambda_{2})\psi_{n-12} + & ... & (a_{n-1n-1} - \lambda_{4})\psi_{n-1n} +\\
+ 2a_{n-1n}\psi_{n1} & + 2a_{n-1n}\psi_{n2} & ... & + 2a_{n-1n}\psi_{nn}\\
       &  &  & \\
   (a_{nn} - \lambda_{1})\psi_{n1} & (a_{nn} - \lambda_{2})\psi_{n2} & ... & (a_{nn} - \lambda_{n})\psi_{nn}\\
\end{array}
\right). $$

\begin{predl}
The minors obtained by the intersection of the first $k<n$ rows or the last $l<n$ rows with any set of $k$ or $l$ columns are semi-invariants with respect to the action of the one-parametric subgroup induced by iso-spectral integral $\frac{1}{2} Tr L^{2}$. The equations of motion of these minors ($M$, $\tilde{M}$) have the following form:
\beq{minors2}
\begin{array}{c}
M^{'}_{\frac{1,2,...,k}{i_{1},i_{2},...,i_{k}}} = (-\sum_{j=1}^{k}a_{jj} + \sum_{i_{m}=i_{1}}^{i_{k}}\lambda_{i_{m}}) M_{\frac{1,2,...,k}{i_{1},i_{2},...,i_{k}}} \ \ ,\\
\tilde{M}^{'}_{\frac{n-l+1,...,n}{i_{1},i_{2},...,i_{l}}} = (\sum_{j=n-l+1}^{n}a_{jj} - \sum_{i_{m}=i_{1}}^{i_{l}}\lambda_{i_{m}}) \tilde{M}_{\frac{n-l+1,...,n}{i_{1},i_{2},...,i_{l}}} \ \ ,
\end{array}
\eq
where in $\frac{a}{b}$ $"a"$ corresponds to the rows, and $"b"$ to the columns.
\end{predl}

$\square$\\
We will prove this proposition by induction. Fixing some $n$, let us consider the minor obtained by the intersection of the first $k$ rows and any set of $k$ columns. For $k=1$ formula (\ref{minors2}) is obvious. Now let us choose some $k$ and assume that for $k-1$ the formula (\ref{minors2}) is true. The $k$-th row decomposition of the minor of the order $k$ has the following form:
\beq{minors-prove}
\begin{array}{c}
M_{\frac{1,2,...,k}{i_{1},i_{2},...,i_{k}}} = \sum_{j=1}^{k}(-1)^{k+j}\psi_{ki_{j}}M_{\frac{1,2,...,k-1}{i_{1},...,i_{\hat{j}},...,i_{k}}}
\end{array}
\eq
After differentiation we get an equation of motion for this minor
\beq{minors-prove2}
\begin{array}{c}
M^{'}_{\frac{1,2,...,k}{i_{1},i_{2},...,i_{k}}} = \sum_{j=1}^{k}(-1)^{k+j}(\psi^{'}_{ki_{j}}M_{\frac{1,2,...,k-1}{i_{1},...,i_{\hat{j}},...,i_{k}}}+\psi_{ki_{j}}M^{'}_{\frac{1,2,...,k-1}{i_{1},...,i_{\hat{j}},...,i_{k}}})\\
\end{array}
\eq
where
\beq{minors-prove3}
\begin{array}{c}
M^{'}_{\frac{1,2,...,k-1}{i_{1},...,i_{\hat{j}},...,i_{k}}}=(-\sum_{j=1}^{k-1}a_{jj} + \sum_{m=1,m \neq j}^{k}\lambda_{i_{m}})M_{\frac{1,2,...,k-1}{i_{1},...,i_{\hat{j}},...,i_{k}}}\\
\psi^{'}_{ki_{j}}=(-a_{kk} + \lambda_{i_{j}})\psi_{ki_{j}}-2(a_{1k}\psi_{1i_{j}} +...+ a_{k-1k}\psi_{k-1i_{j}}).
\end{array}
\eq
Finally we get
\beq{minors-prove4}
\begin{array}{c}
M^{'}_{\frac{1,2,...,k}{i_{1},i_{2},...,i_{k}}} = \sum_{j=1}^{k}(-1)^{k+j}(-\sum_{j=1}^{k}a_{jj} + \sum_{m=1}^{k}\lambda_{i_{m}})\psi_{ki_{j}}M_{\frac{1,2,...,k-1}{i_{1},...,i_{\hat{j}},...,i_{k}}}-\\
-\sum_{j=1}^{k}(-1)^{k+j}2(a_{1k}\psi_{1i_{j}} +...+ a_{k-1k}\psi_{k-1i_{j}})M_{\frac{1,2,...,k-1}{i_{1},...,i_{\hat{j}},...,i_{k}}}.
\end{array}
\eq
Let us consider the second term
\beq{minors-prove5}
\begin{array}{c}
-\sum_{j=1}^{k}(-1)^{k+j}2(a_{1k}\psi_{1i_{j}} +...+ a_{k-1k}\psi_{k-1i_{j}})M_{\frac{1,2,...,k-1}{i_{1},...,i_{\hat{j}},...,i_{k}}}.\\
\end{array}
\eq
The coefficient at every term $-2a_{lk}, \ l=\overline{1,k-1}$ has the following form
\beq{minors-prove6}
\begin{array}{c}
\sum_{j=1}^{k}(-1)^{k+j}\psi_{li_{j}}M_{\frac{1,2,...,k-1}{i_{1},...,i_{\hat{j}},...,i_{k}}}.\\
\end{array}
\eq
This is the minor of the order $k$, obtained from the matrix in which the $l$-th row and the last row are equal. So this minor is equal to zero. And we get from (\ref{minors-prove4})
\beq{minors-prove7}
\begin{array}{c}
M^{'}_{\frac{1,2,...,k}{i_{1},i_{2},...,i_{k}}} = \sum_{j=1}^{k}(-1)^{k+j}(-\sum_{j=1}^{k}a_{jj} + \sum_{m=1}^{k}\lambda_{i_{m}})\psi_{ki_{j}}M_{\frac{1,2,...,k-1}{i_{1},...,i_{\hat{j}},...,i_{k}}}=\\
=(-\sum_{j=1}^{k}a_{jj} + \sum_{m=1}^{k}\lambda_{i_{m}})M_{\frac{1,...,k}{i_{1},...,i_{k}}},
\end{array}
\eq
this is the first formula of (\ref{minors2}). The proof of the second formula of (\ref{minors2}) is similar (use the lower rows instead of upper ones).
$\blacksquare$\\

In a general case for the iso-spectral integral of motion $\frac{1}{d}TrL^{d}, \ d=\overline{2,n} \ $ we get
\beq{psi-d-1}
\Psi^{'} = B_{d} \Psi,
\eq
where $B_{d}$:

\beq{B-d}
B_{d}= (L^{d-1})_{>0} - (L^{d-1})_{<0} =
\left(
\begin{array}{c c c c c c}
 0 & a^{(d-1)}_{12} & a^{(d-1)}_{13} & ... & ... & a^{(d-1)}_{1n}\\
 -a^{(d-1)}_{12} & 0 & a^{(d-1)}_{23} & ... & ... & a^{(d-1)}_{2n}\\
 -a^{(d-1)}_{13} & -a^{(d-1)}_{23} & 0 & ... & ... & a^{(d-1)}_{3n}\\
 ... & ... & ... & ... & ... & ...\\
 ... & ... & ... & ... & ... & a^{(d-1)}_{n-1 n}\\
 -a^{(d-1)}_{1n} & -a^{(d-1)}_{2n} & -a^{(d-1)}_{3n} & ... & -a^{(d-1)}_{n-1 n} & 0\\
\end{array}
\right),
\eq
where $a^{(d-1)}_{ij}$ is matrix element of $L^{d-1}$ which is the product of $d-1$ copies of the Lax operator $L$. It is possible to transform  (\ref{psi-d-1}) to the form, similar to (\ref{N-2}) where instead of $a_{ij}$ there are $a^{(d-1)}_{ij}$ and instead of $\lambda_{j}$ we have $\lambda^{d-1}_{j}$
\beq{psi-d-2}
\begin{array}{c}
\psi^{'}_{ij}=(-a^{(d-1)}_{ii} + \lambda^{d-1}_{j})\psi_{ij}-2\sum^{k=i}_{k=1}a^{(d-1)}_{k-1i}\psi_{k-1j},\\
\ \\
\psi^{'}_{ij}=(a^{(d-1)}_{ii} - \lambda^{d-1}_{j})\psi_{ij}+2\sum^{k=n}_{k=i}a^{(d-1)}_{ik+1}\psi_{k+1j},\\
\end{array}
\eq
So, in general case the following proposition is true:
\begin{predl}
The minors which are obtained by the intersection of the first $k<n$ rows or the last $l<n$ rows with any set of $k$ or $l$ columns are the semi-invariants with respect to the action of the one-parametric subgroup induced by iso-spectral integral $\frac{1}{d} Tr L^{d}, \ d \in \mathbb{N}$. The equations of motion of these minors ($M$, $\tilde{M}$) have the following form:
\beq{minors-d}
\begin{array}{c}
\frac{\partial}{\partial t_{d-1}}M_{\frac{1,2,...,k}{i_{1},i_{2},...,i_{k}}} = (-\sum_{j=1}^{k}a^{(d-1)}_{jj} + \sum_{i_{m}=i_{1}}^{i_{k}}\lambda^{d-1}_{i_{m}}) M_{\frac{1,2,...,k}{i_{1},i_{2},...,i_{k}}} \ \ ,\\
\frac{\partial}{\partial t_{d-1}}\tilde{M}_{\frac{n-l+1,...,n}{i_{1},i_{2},...,i_{l}}} = (\sum_{j=n-l}^{n}a^{(d-1)}_{jj} - \sum_{i_{m}=i_{1}}^{i_{l}}\lambda^{d-1}_{i_{m}}) \tilde{M}_{\frac{n-l+1,...,n}{i_{1},i_{2},...,i_{l}}}.
\end{array}
\eq
\end{predl}

The proof of this Proposition is analogous to the proof of Proposition 3.1.

\section{Families of integrals in involution for the case $n=4$}

In the present section we consider the case $n=4$ and define a method to obtain the additional integrals. This method does not use the chopping procedure. With its help we find in an explicit form all additional integrals of motion in variables $a_{ij}$ and in variables $\lambda, \ \psi$. We also describe two families of integrals in involution.\\

In the papers \cite{EFS, S1, S2} it was discovered that in the case $n=4$ in the Full Kostant-Toda lattice there are two families of integrals in involution. We will show that the same is true for the Full symmetric Toda lattice. Let us give a brief description of the families in involution in the case $n=4$:

\beq{L4-1} L=\left(
\begin{array}{c c c c}
 a_{11} & a_{12} & a_{13} & a_{14}\\
 a_{12} & a_{22} & a_{23} & a_{24}\\
 a_{13} & a_{23} & a_{33} & a_{34}\\
 a_{14} & a_{24} & a_{34} & a_{44}\\
\end{array}
\right).
\eq

It is possible to express the matrix elements via the variables $\lambda, \ \psi$:
\beq{a4}
\begin{array}{c}
a_{ij} = \sum_{k=1}^{4} \lambda_{k}\psi_{ik}\psi_{jk}.
\end{array}
\eq

Using the chopping procedure (\cite{DLNT}) we get the following characteristic polynomials:
\beq{Pk4}
P_{k}(L,\mu) = \sum_{m=0}^{4-2k} E_{m,k}(L)\mu^{4-2k-m}, \,\,\, 0 \leq k \leq 2.
\eq

\beq{Pk4-1}
\begin{array}{c}
P_{0}(L,\mu) = det(L-\mu I),\\
P_{1}(L,\mu) = det(L-\mu I)_{1},\\
\end{array}
\eq

\beq{Pk4-2}
\begin{array}{c}
P_{0}(L,\mu)= \mu^{4} - \mu^{3}TrL + \mu^{2}(\frac{1}{2}(TrL)^{2} - \frac{1}{2}TrL^{2}) - \mu (\frac{1}{3}(TrL)^{3} - \frac{1}{3}TrL^{3}) + detL,\\
\ \\
P_{1}(L,\mu)= \mu^{2}a_{14} + \mu (A_{\frac{34}{13}} + A_{\frac{24}{12}}) + A_{\frac{234}{123}},
\end{array}
\eq

$$E_{0,1} = a_{14}, \,\,\,\, E_{1,1} = A_{\frac{34}{13}} + A_{\frac{24}{12}}, \,\,\,\, E_{2,1} = A_{\frac{234}{123}},\\
$$

The integrals of motion have the following forms:
\beq{Int-n4}
\begin{array}{c}
TrL, \ \ \ \frac{1}{2}TrL^{2}, \ \ \ \frac{1}{3}TrL^{3}, \ \ \ detL,\\

\ \\

I_{1,1}
= \frac{E_{1,1}}{E_{0,1}} = \frac{A_{\frac{34}{13}} + A_{\frac{24}{12}}}{a_{14}}, \ \ \
I_{2,1}= \frac{E_{2,1}}{E_{0,1}} = \frac{A_{\frac{234}{123}}}{a_{14}}.
\end{array}
\eq
where $TrL, \ I_{1,1}$ are Casimirs and the remaining functions are Hamiltonians. These integrals form the first family in involution. Let us consider the isomorphism of the algebras $\rho: \mathfrak{sl}_4(\mathbb{C}) \leftrightarrow \mathfrak{so}_6(\mathbb{C})$. Using the chopping procedure for the Lax operator obtained by the isomorphism $\rho$ we get the second family in involution (see the paper \cite{S1}). In result we get the integral $J$ in addition to the known integrals. This integral in variables $\lambda, \psi$ has the following form:
\beq{J}
J = \frac{A_{\lambda}A^{'}_{\lambda}M_{\frac{12}{12}}M_{\frac{12}{34}} - B_{\lambda}B^{'}_{\lambda} M_{\frac{12}{13}}M_{\frac{12}{24}}}{A_{\lambda}M_{\frac{12}{12}}M_{\frac{12}{34}} - B_{\lambda} M_{\frac{12}{13}}M_{\frac{12}{24}}},
\eq
where
\beq{J-lambda}
\begin{array}{c}

A_{\lambda} = (\lambda_{1} - \lambda_{3})(\lambda_{2} - \lambda_{4}),\\

A^{'}_{\lambda}= (\lambda_{1} + \lambda_{3})(\lambda_{2} + \lambda_{4}),\\

B_{\lambda}= (\lambda_{1} - \lambda_{2})(\lambda_{3} - \lambda_{4}),\\

B^{'}_{\lambda}= (\lambda_{1} + \lambda_{2})(\lambda_{3} + \lambda_{4}).

\end{array}
\eq

The second family in involution are formed by the integrals $TrL, \ \frac{1}{2}TrL^{2}, \ \frac{1}{3}TrL^{3}, \ detL, \ \ I_{1,1}$ and the integral $J$. This integral does not commute with integral $I_{2,1}$.

It turns out that it is possible to obtain the family in involution by another method which does not use the chopping procedure and isomorphism $\rho: \mathfrak{sl}_4(\mathbb{C}) \leftrightarrow \mathfrak{so}_6(\mathbb{C})$; which is important since this isomorphism does not exist in case $n>4$, when there is only homomorphism $\mathfrak{sl}_n(\mathbb{C}) \rightarrow \mathfrak{so}_C^{m}(\mathbb{C})$.

Let us consider the matrix $L^{k} = \underbrace{L \cdot L \cdot ... \cdot L}_{k}, \ k \geq 2$, where $L$ is the Lax operator $L, \ n=4$. Let us denote
the matrix elements of the matrix $L^{k}$ by $a^{(k)}_{ij}$. Let us make the following observation: the element $a_{14}$ is a semi-invariant, moreover the element $a^{(k)}_{14}$\footnote[1]{D. Talalaev pointed out this property of matrix element $a^{(k)}_{14}$. The proof was based on the consideration of the action of subgroup $b_{+}$ on $a_{14}$.} is a semi-invariant. Indeed, if we express the element $a_{14}$ via the variables $\lambda$ and $\psi$
$$
a_{14} = \lambda_{1}\psi_{11}\psi_{41} + \lambda_{2}\psi_{12}\psi_{42} + \lambda_{3}\psi_{13}\psi_{43} + \lambda_{4}\psi_{14}\psi_{44}.$$
One can see from this expression that the dynamics of the element $a_{14}$ with respect to the action of one-parametric subgroup induced by the iso-spectral integral of motion $\frac{1}{2}TrL^{2}$ (see (\ref{minors2})) has the following form:
\beq{Dyna14}
a^{'}_{14}= (a_{44}-a_{11}) a_{14}.
\eq
This means that the element $a_{14}$ is the semi-invariant. Further notice that from
$$L \Psi = \Psi \Lambda,$$
it follows that
$$L^{k} \Psi= \Psi \Lambda^{k},$$
where $\Psi, \ \Lambda$ are the eigenvector and eigenvalue matrices. Then the element $a^{(k)}_{14}$ has the following form:
$$
 a^{(k)}_{14} = \lambda^{k}_{1}\psi_{11}\psi_{41} + \lambda^{k}_{2}\psi_{12}\psi_{42} + \lambda^{k}_{3}\psi_{13}\psi_{43} + \lambda^{k}_{4}\psi_{14}\psi_{44}.$$
Its dynamics is given by (\ref{minors2}):
\beq{Dyna15}
 (a^{(k)}_{14})^{'} = (a_{44}-a_{11}) a^{(k)}_{14}.
\eq
It is obvious that the ratio of $a^{(k_{1})}_{14}$ to $a^{(k_{2})}_{14}$ is the integral of motion where $k_{1}, \ k_{2} \in \mathbb{N}$.
Now the question is if there exist other minors $A$ of the Lax which are semi-invariants and have the dynamics like (\ref{Dyna15}):
\beq{Dyna16}
 A^{'} = f(a) A,
\eq
where $f(a)$ is some function of the matrix elements of the Lax. It turns out that such minors exist.
Let us consider the minor $A_{\frac{34}{12}}$. The expression of this minor in the terms of the variables $\lambda$ and $\psi$ has the following form:
\beq{A3412n4}
A_{\frac{34}{12}}=(\lambda_{1}\lambda_{2} + \lambda_{3}\lambda_{4})M_{\frac{12}{34}}M_{\frac{12}{12}} +
(-\lambda_{1}\lambda_{3} - \lambda_{2}\lambda_{4})M_{\frac{12}{13}}M_{\frac{12}{24}} + (\lambda_{1}\lambda_{4} + \lambda_{2}\lambda_{3})M_{\frac{12}{23}}M_{\frac{12}{14}}.
\eq
The dynamics of the minor $M_{\frac{12}{ij}}$, due to (\ref{minors2}), has the following form:
$$M^{'}_{\frac{12}{ij}} = (-a_{11}-a_{22}+\lambda_{i}+\lambda_{j})M_{\frac{12}{ij}},$$
So for $A_{\frac{34}{12}}$ we get:
\beq{Din-A3412n4}
A^{'}_{\frac{34}{12}} = (-2(a_{11}+a_{22}) + TrL)A_{\frac{34}{12}}.
\eq
and for minors $A^{k}_{\frac{34}{12}}$ we get:
\beq{DinA3412Kn4}
\begin{array}{c}
A^{(k)}_{\frac{34}{12}}=(\lambda^{k}_{1}\lambda^{k}_{2} + \lambda^{k}_{3}\lambda^{k}_{4})M_{\frac{12}{34}}M_{\frac{12}{12}} +
(-\lambda^{k}_{1}\lambda^{k}_{3} - \lambda^{k}_{2}\lambda^{k}_{4})M_{\frac{12}{13}}M_{\frac{12}{24}} + (\lambda^{k}_{1}\lambda^{k}_{4} + \lambda^{k}_{2}\lambda^{k}_{3})M_{\frac{12}{23}}M_{\frac{12}{14}},\\

\ \\

(A^{(k)}_{\frac{34}{12}})^{'}= (-2(a_{11}+a_{22}) + TrL)A^{k}_{\frac{34}{12}}
\end{array}
\eq

\paragraph{The integrals of motion}
Using the above mentioned approach let us define the integrals of motion.
\beq{Integrals-n4}
\begin{array}{c}
TrL, \ \ \ \frac{1}{2}TrL^{2}, \ \ \ \frac{1}{3}TrL^{3}, \ \ \ detL,\\
\end{array}
\eq

$$I_{1,1}=\frac{a^{(2)}_{14}}{a_{14}}
= \frac{\lambda^{2}_{1}\psi_{11}\psi_{41} + \lambda^{2}_{2}\psi_{12}\psi_{42} + \lambda^{2}_{3}\psi_{13}\psi_{43} + \lambda^{2}_{4}\psi_{14}\psi_{44}}{\lambda_{1}\psi_{11}\psi_{41} + \lambda_{2}\psi_{12}\psi_{42} + \lambda_{3}\psi_{13}\psi_{43} + \lambda_{4}\psi_{14}\psi_{44}},$$

$$\tilde{I}_{2,1}= -\frac{\lambda^{-1}_{1}\psi_{11}\psi_{41} + \lambda^{-1}_{2}\psi_{12}\psi_{42} + \lambda^{-1}_{3}\psi_{13}\psi_{43} + \lambda^{-1}_{4}\psi_{14}\psi_{44}}{\lambda_{1}\psi_{11}\psi_{41} + \lambda_{2}\psi_{12}\psi_{42} + \lambda_{3}\psi_{13}\psi_{43} + \lambda_{4}\psi_{14}\psi_{44}} = \frac{I_{2,1}}{detL},$$
instead of this integral we can consider
$$I=\frac{a^{(3)}_{14}}{a_{14}}= \frac{\lambda^{3}_{1}\psi_{11}\psi_{41} + \lambda^{3}_{2}\psi_{12}\psi_{42} + \lambda^{3}_{3}\psi_{13}\psi_{43} + \lambda^{3}_{4}\psi_{14}\psi_{44}}{\lambda_{1}\psi_{11}\psi_{41} + \lambda_{2}\psi_{12}\psi_{42} + \lambda_{3}\psi_{13}\psi_{43} + \lambda_{4}\psi_{14}\psi_{44}},$$
and one more integral which does not commute with $I_{2,1}$ and $I$
$$J =\frac{A^{(2)}_{\frac{34}{12}}}{A_{\frac{34}{12}}}= -\frac{(\lambda^{2}_{1}\lambda^{2}_{2} + \lambda^{2}_{3}\lambda^{2}_{4})M_{\frac{12}{34}}M_{\frac{12}{12}} +
(-\lambda^{2}_{1}\lambda^{2}_{3} - \lambda^{2}_{2}\lambda^{2}_{4})M_{\frac{12}{13}}M_{\frac{12}{24}} + (\lambda^{2}_{1}\lambda^{2}_{4} + \lambda^{2}_{2}\lambda^{2}_{3})M_{\frac{12}{23}}M_{\frac{12}{14}}}{(\lambda_{1}\lambda_{2} + \lambda_{3}\lambda_{4})M_{\frac{12}{34}}M_{\frac{12}{12}} +
(-\lambda_{1}\lambda_{3} - \lambda_{2}\lambda_{4})M_{\frac{12}{13}}M_{\frac{12}{24}} + (\lambda_{1}\lambda_{4} + \lambda_{2}\lambda_{3})M_{\frac{12}{23}}M_{\frac{12}{14}}}.$$
Observe that taking into account the Pl\"ucker relation in the case $n=4$ for $Gr_{2}(4)$
\beq{Plucker-n4}
M_{\frac{12}{34}}M_{\frac{12}{12}}- M_{\frac{12}{13}}M_{\frac{12}{24}} + M_{\frac{12}{23}}M_{\frac{12}{14}} = 0,
\eq
it is possible to get the integral $J$ in the form of (\ref{J}).

\begin{figure}[!t]
\begin{center}
\includegraphics[width=250pt,height=270pt]{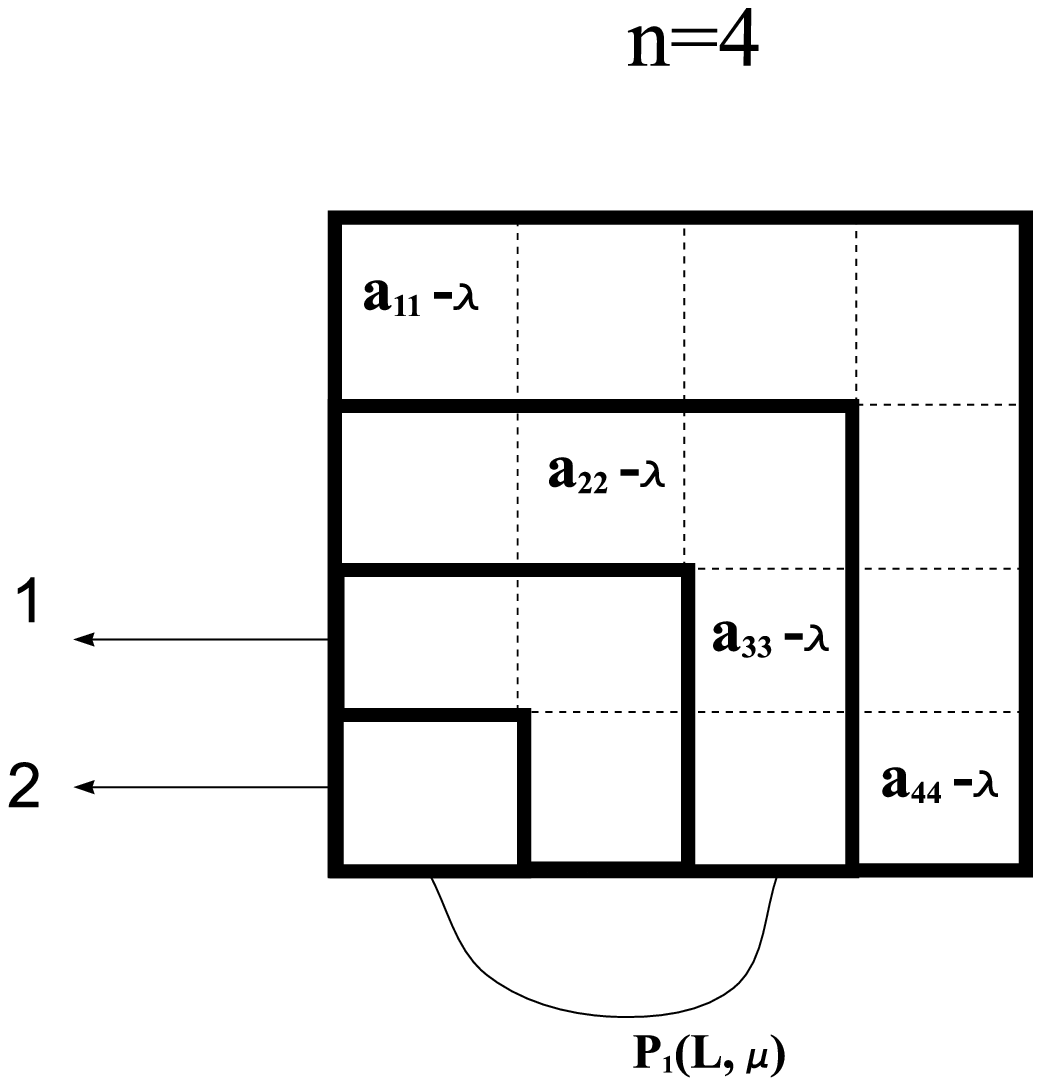}
\end{center}
\end{figure}

There is the case $n=4$ on the picture. The curve below points out the characteristic polynomial $P_{1}(L,\mu)$ in the case $n=4$. This polynomial is obtained by the chopping procedure. The curve connects the leading coefficient -- $a_{14}$ -- at $\mu^{2}$, on which the first family in involution is constructed and the boundary element of the minor obtained by chopping procedure. The upper arrow points out the numeral $1$ which corresponds to an additional integrals which obtained from the minors $A^{k}_{\frac{34}{12}}$ without the chopping procedure. The arrow going from the minor $a_{14}$ points out the numeral $2$. It corresponds to the two integrals constructed from the minors $a^{k}_{14}$.

\section{Structure of the integrals of motion. Families of integrals}

In this section we generalize our method to obtain the additional integrals of motion and explore the structure of integrals and families of integrals.\\

\subsection{General structure of the full non-commutative family of integrals of motion}
The number of independent non-involutive integrals without taking into account the Casimirs \cite{BG-1, GS}
\beq{Integrals-number}
\begin{array}{c}
N_{n} = \frac{n(n-1)}{2} - [\frac{n+1}{2}] + 1
\end{array}
\eq
consists of three different contributions
\beq{IntContr}
\begin{array}{c}
N_{n} \equiv N_{n}^{Iso} + N_{n}^{Chopp} + N_{n}^{Add}
\end{array}
\eq
which come from the iso-spectral integrals $N_{n}^{Iso}$, the integrals $N_{n}^{Chopp}$ derived by the chopping procedure with exclusion of the Casimirs, and the additional integrals $N_{n}^{Add}$ which extend the above--mentioned two families to the full set of the integrals, respectively. The sum of the iso-spectral integrals and the integrals obtained by the chopping procedure, which ensures the Liouville integrability, is \cite{DLNT}
\beq{IntChop}
\begin{array}{c}
N_{n}^{Iso} + N_{n}^{Chopp}=[\frac{n^{2}}{4}] \ .
\end{array}
\eq
From the equalities (\ref{Integrals-number}), (\ref{IntContr}) and (\ref{IntChop}) one can obviously conclude that
\beq{IntChopAddEq}
\begin{array}{c}
N_{n}^{Add}=N_{n}^{Chopp} \equiv [\frac{(n-2)^{2}}{4}].
\end{array}
\eq

\subsection{Semi-invariants and Integrals}
In this subsection we explore the structure of the semi-invariants of motion and show that the semi-invariants have the quadratic dependence on the minors of the eigenvector matrix of the Lax operator. Also we define the integrals of motion.

\begin{predl}
The minors $ \ \ A^{(k)}_{\frac{n-m+1,...,n}{1,2,...,m}}, \ \ \ n>m \ \ $ of the matrices $L^{k}$ are semi-invariants with respect to the action of one-parametric subgroup induced by the iso-spectral integrals of motion. These minors take the following form in variables $\lambda, \ \psi$:
\beq{A-lambda-psi}
A^{(k)}_{\frac{n-m+1,...,n}{1,2,...,m}}=\sum_{i_{1},i_{2},...,i_{m}} \lambda^{k}_{i_{1}}\lambda^{k}_{i_{2}} \cdot...\cdot \lambda^{k}_{i_{m}} M_{\frac{1,2,...,m}{i_{1},i_{2},...,i_{m}}}M_{\frac{n-m+1,...,n}{i_{1},i_{2},...,i_{m}}}.\\
\eq
The integrals of motion equal to the combinations of these minors have the following form:
\beq{A-lambda-psi-Integrals}
J_{k_{1},k_{2}}= \frac{A^{(k_{1})}_{\frac{n-m+1,...,n}{1,2,...,m}}}{A^{(k_{2})}_{\frac{n-m+1,...,n}{1,2,...,m}}}.
\eq
\end{predl}

$\square$\\
First of all we proof that there are no products like $\lambda^{2}_{l}$ in the decomposition of the minor $A_{\frac{n-m+1,...,n}{1,2,...,m}}$
(left lower angle minor $m \times m$) of the Lax operator in variables $\lambda, \ \psi$. By the definition the minor $A_{\frac{n-m+1,...,n}{1,2,...,m}}$ has the following decomposition in terms of the matrix elements of the Lax operator
\beq{minordet}
A_{\frac{n-m+1,...,n}{1,2,...,m}} = \sum_{p}(-1)^{t(p)}a_{n-m+1,j_{1}} \cdot a_{n-m+2,j_{2}} \cdot ... \cdot a_{n,j_{m}},
\eq
where $t(p)$ is the number of transpositions in the permutation $p$ - $(j_{1}j_{2}...j_{m})$ - of the set $\{1,2,...,m\}$. Let us call by "path" any product of $a_{ij}$ which is one of the $m!$ terms in the decomposition of the minor $A_{\frac{n-m+1,...,n}{1,2,...,m}}$. Fix $l$ and assume that $\lambda^{2}_{l}$ appears in the decomposition of some path in variables $\lambda, \ \psi$. For the sake of definiteness we will assume that $\lambda^{2}_{l}$ arises from two elements $a_{ij}$:
\beq{aanda}
\begin{array}{c}
a_{i_{1}j_{a}}=\sum_{l_{1}=1}^{n} \lambda_{l_{1}} \psi_{i_{1}l_{1}} \psi_{j_{a}l_{1}},\\
a_{i_{2}j_{b}}=\sum_{l_{2}=1}^{n} \lambda_{l_{2}} \psi_{i_{2}l_{2}} \psi_{j_{b}l_{2}}.
\end{array}
\eq
Extracting the terms with $\lambda_{l}$ from the formulas for $a_{ij}$ we get $\lambda_{l} \psi_{i_{1}l} \psi_{j_{a}l}$ and $\lambda_{l} \psi_{i_{2}l} \psi_{j_{b}l}$. Their product is $c_{1} = \lambda^{2}_{l} \psi_{i_{1}l} \psi_{j_{a}l} \psi_{i_{2}l} \psi_{j_{b}l}$. Now let us consider another path (we will call it "dual path") which is distinguished from the given path by one transposition -- $j_{1}j_{2}...j_{b}...j_{a}...j_{m}$ instead of $j_{1}j_{2}...j_{a}...j_{b}...j_{m}$ and in consequence by the factor  -- (-1). Then the dual path is distinguished from the given path by only two matrix elements $a_{ij}$ (and by the factor (-1)):
\beq{aanda2}
\begin{array}{c}
a_{i_{1}j_{b}}=\sum_{l_{3}=1}^{n} \lambda_{l_{3}} \psi_{i_{1}l_{3}} \psi_{j_{b}l_{3}},\\
a_{i_{2}j_{a}}=\sum_{l_{4}=1}^{n} \lambda_{l_{4}} \psi_{i_{2}l_{2}} \psi_{j_{2}l_{2}}.
\end{array}
\eq
Extracting the terms with $\lambda_{l}$ from the formulas for $a_{ij}$ we get $\lambda_{l} \psi_{i_{1}l} \psi_{j_{b}l}$ and $\lambda_{l} \psi_{i_{2}l} \psi_{j_{a}l}$. Their product is $c_{2} = \lambda^{2}_{l} \psi_{i_{1}l} \psi_{j_{b}l} \psi_{i_{2}l} \psi_{j_{a}l}$. As we see the products $c_{1}$ and $c_{2}$ coincide except for the factor $(-1)$. The coefficients at $c_{1}$ and $c_{2}$ consist of products $a_{ij}$ and as it was mentioned above coincide. So, the sum $c_{1}+c_{2}$ is equal to zero and the term with $\lambda^{2}_{l}$ vanishes. It is obvious that in the decomposition of minor $A_{\frac{n-m+1,...,n}{1,2,...,m}}$ in the term of $\lambda$ and $\psi$ there can not appear any degree of $\lambda_{l}$ except the one.\\

Now let us show that the decomposition of left lower angle minor of $L$ consists of the sum with the terms having the following form $\lambda_{i_{1}}\lambda_{i_{2}} \cdot...\cdot \lambda_{i_{m}} M_{\frac{1,2,...,m}{i_{1},i_{2},...,i_{m}}}M_{\frac{n-m+1,...,n}{i_{1},i_{2},...,i_{m}}}$. The number of the terms in this decomposition is determined by the number ($n_{\lambda}$) of the different products $\lambda_{i_{1}}\lambda_{i_{2}} \cdot...\cdot \lambda_{i_{m}}$. This number is equal to $n_{\lambda}=\frac{n!}{k!(n-m)!}$. Choosing some product of $\lambda$ which consists of the sequence of the products of the different $\lambda$ let us find what is the product of the elements $\psi$ as the coefficient at the chosen sequence.

For the sake of simplicity let us chose a sequence $Seq_{1}=\lambda_{1}\lambda_{2} \cdot...\cdot \lambda_{m}$ and fix some path $P_{1}$ $(-1)^{t_{1}(p)}a_{n-m+1,j_{1}} \cdot a_{n-m+2,j_{2}} \cdot ... \cdot a_{n,j_{m}}$. So we fixed the set of $j_{l}$. Because $a_{ij}=\sum_{k=1}^{n} \lambda_{k} \psi_{ik} \psi_{jk}$ let us consider the element $\psi_{n-m+s,s},$ at $\lambda_{s}, s=\overline{1,m}$. Then there is $\psi_{j_{s}s}$ at $\lambda_{s}\psi_{n-m+s,s}$. So, from the path $P_{1}$ we get the product $(-1)^{t_{1}(p)}\psi_{j_{1}1}\psi_{j_{2}2}...\psi_{j_{m}m}$ at $\prod \lambda_{s}\psi_{n-m+s,s}$. For the other path $P_{2}$ we get the product $(-1)^{t_{2}(p)}\prod \psi_{j_{l}s}, \ l=\overline{1,m}$ at $\prod \lambda_{s}\psi_{n-m+s,s}$ with the factor $(-1)^{t_{2}(p)}$ which is inherited from the path $P_{2}$. After same manipulations for all permutations $j_{l}$ (for all paths), we get the coefficient at $\prod \lambda_{s}\psi_{n-m+s,s}$ which is equal to the sum $\sum_{p}(-1)^{t(p)}\psi_{j_{1}1}\psi_{j_{2}2}...\psi_{j_{m}m}$. This sum is the minor $M_{\frac{1,2...m}{1,2...m}}$ of the matrix $\Psi$. If we consider the permutation of the indices $s$, which means choosing of other elements $\psi$ from the given path with the same set of $\lambda_{s}$, we get the same minor $M_{\frac{1,2...m}{1,2...m}}$ multiplied by the factor $(-1)^{t_{a}(p)}$, where $t_{a}(p)$ is the number of transposition of the set $\{ 1,2...m \}$, the indices at $\lambda_{s}$. This is clear since the same permutation is the permutation of indexes ${j_{s}}$. Now we can factor out the term $\lambda_{1}\lambda_{2} \cdot...\cdot \lambda_{m}M_{\frac{1,2...m}{1,2...m}}$. We get the sum of $m!$ elements of the following form $(-1)^{t(p)}\psi_{n-m+1,1}\psi_{n-m+2,2}...\psi_{n,m}$ which is the minor $M_{\frac{n-m+1,n-m+2...n}{1,2...m}}$ of the matrix $\Psi$. After the same calculations for the other sequences of $\lambda$, we get the formula
(\ref{A-lambda-psi}) for $k=1$.

It remains to show that any term of the decomposition (\ref{minordet}) belongs to some expression of the form $\lambda_{i_{1}}\lambda_{i_{2}} \cdot...\cdot \lambda_{i_{m}} M_{\frac{1,2,...,m}{i_{1},i_{2},...,i_{m}}}M_{\frac{n-m+1,...,n}{i_{1},i_{2},...,i_{m}}}$. Indeed, from the above arguments it follows that the products of minors contain all possible sequences of the products of the elements $\psi$ of two types. Any term of the decomposition consists just of two products of such sequences which multiplied by the product of $\lambda$. It follows directly from the decomposition of the products of $a_{ij}=\sum_{k=1}^{n} \lambda_{k} \psi_{ik} \psi_{jk}$.

Formula (\ref{A-lambda-psi}) for each $k$ follows from $L^{k}\Psi = \Psi \Lambda^{k}$, and formula (\ref{A-lambda-psi-Integrals}) follows from the Propositions of the Section $3$.
$\blacksquare$\\

\subsection{Integrals of motion obtained by the chopping procedure and Pl\"ucker coordinates}

In this subsection we give the formula which expresses the additional integrals of motion obtained by the chopping procedure via the Pl\"ucker coordinates on the flag space.

\begin{predl}
The involutive integrals (\ref{Integrals}) derived by the chopping procedure can be represented as
$$
I_{m,k} = \frac{\sum_{i_{1}<...<i_{s}<...<i_{k}} \lambda_{i_{1}} \cdot ... \cdot \lambda_{i_{k}} \sum_{j_{1}<...<j_{m}, \ j_{t} \neq i_{s}} \lambda_{j_{1}} \cdot ... \cdot \lambda_{j_{m}} M_{\frac{1,...,k}{i_{1},...,i_{k}}}M_{\frac{n-k+1,...,n}{i_{1},...,i_{k}}}}{\sum_{l_{1}<...<l_{r}<...l_{k}} \lambda_{l_{1}} \cdot...\cdot \lambda_{l_{k}} M_{\frac{1,...,k}{l_{1},...,l_{k}}}M_{\frac{n-k+1,...,n}{l_{1},...,l_{k}}}},$$
\beq{Integrals-Plukk}
\begin{array}{c}
1 \leq s,r \leq k, \ 1 \leq t \leq m, \ 1 \leq i_{s},j_{t},l_{r} \leq n\\
\ \\
0 \leq k \leq [\frac{n-1}{2}],  \ \ \ 1 \leq m \leq n-2k.
\end{array}
\eq
\end{predl}

$\square$\\

First of all we will determine what minors constitute the coefficient $E_{m,k}$. Let us consider the chopping procedure and fix $k$ and $m$. Then we have (see (\ref{Pkn}))
\beq{ChoppK}
\begin{array}{c}
\det(L - \mu I)_{k} =  E_{0,k} \mu^{n-2k} +...+E_{m,k} \mu^{n-2k-m}+...+E_{n-2k,k}=0 ,
\end{array}
\eq
The product $E_{m,k} \cdot \mu^{n-2k-m}$ consists of the minors of $(n-k)\times(n-k)$ matrix $(L - \mu I)_{k}$ in such a way that there are $n-2k-m$ elements $\mu$ in this product. The elements $\mu$ are situated on the main diagonal of matrix $(L - \mu I)$ so the coefficient $E_{m,k}$ consists of all minors $(m+k)\times(m+k)$ formed by deleting $n-2k-m$ rows and $n-2k-m$ columns with the same indices as the rows. The index values of deleted rows and columns are limited to $k$ and $n-k+1 $ from above and below, respectively. And we get the formula for the coefficient $E_{m,k}$
\beq{EmkA}
\begin{array}{c}
E_{m,k} = \sum_{i_{k+1}<...<i_{k+m}} A_{\frac{i_{k+1},...,i_{k+m},n-k+1,...,n}{1,...,k,i_{k+1},...,i_{k+m}}},\\
k<i_{k+1}<...<i_{k+m}<n-k+1.
\end{array}
\eq
The number of the terms $A$ is equal to $\frac{(n-2k)!}{(n-2k-m)! \ m!}$. All minors in (\ref{EmkA}) have the following form in variables $\lambda, \ \psi$:
\beq{EmkA-2}
\begin{array}{c}
A_{\frac{i_{k+1},...,i_{k+m},n-k+1,...,n}{1,...,k,i_{k+1},...,i_{k+m}}}=\sum_{j_{1}<...<j_{s}<...<j_{k+m}} \lambda_{j_{1}} \cdot ... \cdot \lambda_{j_{k+m}} M_{\frac{1,...,k,i_{k+1},...,i_{k+m}}{j_{1},...,j_{k+m}}}M_{\frac{i_{k+1},...,i_{k+m},n-k+1,...,n}{j_{1},...,j_{k+m}}},\\
1 \leq j_{s} \leq n,
\end{array}
\eq
This formula is the same as (\ref{A-lambda-psi}) for $A_{\frac{n-m+1,...,n}{1,2,...,m}}$ but without some $n-2k-m$ rows and $n-2k-m$ columns with the same indices as the rows. The proof of (\ref{EmkA-2}) is the same as for (\ref{A-lambda-psi}).

Let us fix some chosen set of $\lambda_{j_{s}}$ -- $\{\lambda_{a_{1}},..., \lambda_{a_{s}},..., \lambda_{a_{k+m}}\}$ -- the same for each minors $A$ in (\ref{EmkA}) after decomposition (\ref{EmkA-2}). Summing these terms we get
\beq{EmkA-3}
\begin{array}{c}
\lambda_{a_{1}} \cdot ...\cdot \lambda_{a_{s}} \cdot... \cdot \lambda_{a_{k+m}}  \sum_{i_{k+1}<...<i_{k+m}}  M_{\frac{1,...,k,i_{k+1},...,i_{k+m}}{a_{1},...,a_{k+m}}}M_{\frac{i_{k+1},...,i_{k+m},n-k+1,...,n}{a_{1},...,a_{k+m}}},
\end{array}
\eq
where for fixed indices we use the notation $a_{s}$.
Decomposing one of the minors in (\ref{EmkA-3}) -- $M_{\frac{1,...,k,i_{k+1},...,i_{k+m}}{a_{1},...,a_{k+m}}}$ -- into the sum of the minor products of matrices $m \times m$ and $k \times k$ (Laplace expansion) we get:
\beq{EmkA-4}
\begin{array}{c}
M_{\frac{1,...,k,i_{k+1},...,i_{k+m}}{a_{1},...,a_{s},...,a_{k+m}}}=\sum_{l_{1}<...<l_{r}<...<l_{k}}  M_{\frac{1,...,k}{l_{1},...,l_{k}}}(-1)^{p+q}M_{\frac{i_{1},...,i_{m}}{l_{k+1},...,l_{k+m}}},\\
l_{r} \in a_{s}, \ a=1+...+k, \ b=l_{1}+...+l_{k}.
\end{array}
\eq
In other words in (\ref{EmkA-4}) the sum is taken over all permutations of $k$ columns with the indices from the set $\{ a_{1},...,a_{k+m} \}$. So we have from (\ref{EmkA-3})
\beq{EmkA-5}
\begin{array}{c}
\sum_{i_{k+1}<...<i_{k+m}} (\sum_{l_{1}<...<l_{r}<...<l_{k}}  M_{\frac{1,...,k}{l_{1},...,l_{k}}}(-1)^{p+q}M_{\frac{i_{k+1},...,i_{k+m}}{l_{k+1},...,i_{k+m}}})
M_{\frac{i_{k+1},...,i_{k+m},n-k+1,...,n}{a_{1},...,a_{k+m}}}\\
=\sum_{l_{1}<...<l_{r}<...<l_{k}}  M_{\frac{1,...,k}{l_{1},...,l_{k}}}(-1)^{p+q} \sum_{i_{k+1}<...<i_{k+m}} M_{\frac{i_{k+1},...,i_{k+m}}{l_{k+1},...,l_{k+m}}}M_{\frac{i_{k+1},...,i_{k+m},n-k+1,...,n}{a_{1},...,a_{k+m}}}.
\end{array}
\eq
Now we fix indices $\{ l_{1},...,l_{k} \}$ and transform the minor $M_{\frac{i_{k+1},...,i_{k+m},n-k+1,...,n}{a_{1},...,a_{k+m}}}$ into cofactor (remind that $\Psi \in SO(n,\mathbb{R})$):
\beq{EmkA-6}
\begin{array}{c}
M_{\frac{i_{k+1},...,i_{k+m},n-k+1,...,n}{a_{1},...,a_{k+m}}}=(-1)^{p_{1}+q_{1}}
М_{\frac{1,...,k,\tilde{i}_{k+1},...,\tilde{i}_{n-k-m}}{\tilde{a}_{1},...,\tilde{a}_{n-k-m}}}=
C_{\frac{i_{k+1},...,i_{k+m},n-k+1,...,n}{a_{1},...,a_{k+m}}}\\
p_{1}=1+...+ k + \tilde{i}_{k+1}+...+\tilde{i}_{n-k-m}, \ q_{1}=\tilde{a}_{1}+...+\tilde{a}_{n-k-m}.
\end{array}
\eq
Now let us add to the sum $\sum_{i_{k+1}<...<i_{k+m}}(...)$ in (\ref{EmkA-5}) the products of minors of matrix $(n-k) \times (n-k)$ with all sequences of $m$ indices $i_{z}$ from the set $\{ 1,...,n-k \}$ which are not in the sum $\sum_{i_{k+1}<...<i_{k+m}}(...)$. Note that in each sequences of $m$ indices $i_{z}$ there is at least one index which belongs to the set $\{ 1,...,k \}$, we get:
\beq{EmkA-7}
\begin{array}{c}
M_{\frac{1,...,k}{l_{1},...,l_{k}}}(-1)^{p+q} \times\\
\times (\sum_{i_{k+1}<...<i_{k+m}} M_{\frac{i_{k+1},...,i_{k+m}}{l_{k+1},...,l_{k+m}}}C_{\frac{i_{k+1},...,i_{k+m},n-k+1,...,n}{a_{1},...,a_{k+m}}}
+\sum_{i_{z_{1}}<...<i_{z_{m}}}M_{\frac{i_{z_{1}},...,i_{z_{m}}}{l_{k+1},...,l_{k+m}}}
C_{\frac{i_{z_{1}},...,i_{z_{m}}}{a_{1},...,a_{k+m}}}).
\end{array}
\eq
and
\beq{EmkA-8}
\begin{array}{c}
(-1)^{p+q}\times (\sum_{i_{k+1}<...<i_{k+m}} M_{\frac{i_{k+1},...,i_{k+m}}{l_{k+1},...,l_{k+m}}}C_{\frac{i_{k+1},...,i_{k+m},n-k+1,...,n}{a_{1},...,a_{k+m}}}
+\sum_{i_{z_{1}}<...<i_{z_{m}}}M_{\frac{i_{z_{1}},...,i_{z_{m}}}{l_{k+1},...,l_{k+m}}}
C_{\frac{i_{z_{1}},...,i_{z_{m}}}{a_{1},...,a_{k+m}}})\\
=M_{\frac{n-k+1,...,n}{l_{1},...,l_{k}}}.
\end{array}
\eq
We do the transformation (\ref{EmkA-6}) for all $M_{\frac{1,...,k}{l_{1},...,l_{k}}}$. The sum of the additional terms are equal to zero because
\beq{EmkA-9}
\begin{array}{c}
\sum_{l_{1}<...<l_{k}}
M_{\frac{1,...,k}{l_{1},...,l_{k}}}(-1)^{p+q}
\sum_{i_{z_{1}}<...<i_{z_{m}}}M_{\frac{i_{z_{1}},...,i_{z_{m}}}{l_{k+1},...,l_{k+m}}}
C_{\frac{i_{z_{1}},...,i_{z_{m}}}{a_{1},...,a_{k+m}}}\\
=\sum_{i_{z_{1}}<...<i_{z_{m}}}C_{\frac{i_{z_{1}},...,i_{z_{m}}}{a_{1},...,a_{k+m}}}
\sum_{l_{1}<...<l_{k}}M_{\frac{1,...,k}{l_{1},...,l_{k}}}(-1)^{p+q}
M_{\frac{i_{z_{1}},...,i_{z_{m}}}{l_{k+1},...,l_{k+m}}}
\end{array}
\eq
and in the minor $\sum_{l_{1}<...<l_{k}}M_{\frac{1,...,k}{l_{1},...,l_{k}}}(-1)^{p+q}
M_{\frac{i_{z_{1}},...,i_{z_{m}}}{l_{k+1},...,l_{k+m}}}$ there are at least two equal rows.

So we get
\beq{EmkA-10}
\begin{array}{c}
\lambda_{j_{1}} \cdot ...\cdot \lambda_{j_{k+m}} \sum_{i_{k+1}<...<i_{k+m}} M_{\frac{1,...,k,i_{k+1},...,i_{k+m}}{j_{1},...,j_{k+m}}}M_{\frac{i_{k+1},...,i_{k+m},n-k+1,...,n}{j_{1},...,j_{k+m}}} \\
=\lambda_{j_{1}} \cdot ...\cdot \lambda_{j_{k+m}} \sum_{l_{1}<...<l_{k}} M_{\frac{1,...,k}{l_{1},...,l_{k}}}M_{\frac{n-k+1,...,n}{l_{1},...,l_{k}}}
\end{array}
\eq
and finally we get the formula (\ref{Integrals-Plukk}) after the change of notation and summation.

$\blacksquare$\\

Note that it is possible to express the integrals (\ref{Integrals-Plukk}) via the integrals (\ref{A-lambda-psi-Integrals}) using eqs. (\ref{A-lambda-psi}), which allow to express all products $M_{\frac{1,2,...,k}{i_{1},i_{2},...,i_{k}}}M_{\frac{n-k+1,...,n}{i_{1},i_{2},...,i_{k}}}$ entering into eq. (\ref{Integrals-Plukk}) in terms of  $A^{(k)}_{\frac{n-m+1,...,n}{1,2,...,m}}$.

\subsection{Integrals of motion in variables $\lambda$ and $\psi$}
In this subsection we show that the set of the integrals of the form (\ref{A-lambda-psi-Integrals}) constructed from minors of the eigenvector matrix of the Lax operator does in fact determine the full non-commutative family of the integrals of the full symmetric Toda lattice.\\

In the previous subsection we saw that one can express the integrals of the full symmetric Toda lattice in the terms of $\lambda$ and $M(\psi)$.
Also we know that $M(\psi)$ is semi-invariant.
The question is: how many independent integrals of motion in variables $\psi$ there exist?
In papers \cite{BG-1},\cite{GS} the formula for the number of independent integrals of motion using the Lie-algebraic approach was concocted out.
We will get the same formula using the facts about the flag space.

It is known that using the Pl\"ucker embedding we can describe the flag space.
\beq{map-fl}
\begin{array}{c}
FL_{n}(\mathbb{R}) \hookrightarrow \mathbb{RP}^{n-1} \times ... \times \mathbb{RP}^{C^{n}_{k_{1}}-1} \times ... \times (\mathbb{RP}^{C^{n}_{k_{2}}-1})^{\ast} \times ... \times (\mathbb{RP}^{n-1})^{\ast},\\
\ \\
1 \leq k_{1} \leq [\frac{n}{2}] < k_{2} \leq n-1.
\end{array}
\eq
Let us consider a basis $e_{i}$ on the vector space $V^{n}= \mathbb{R}^{n}$ and define the basis $e^{\ast}_{i}$ of the dual projective space via unit volume:
\beq{fl-1}
\begin{array}{c}
e_{i} \wedge e^{\ast}_{i} = e_{1} \wedge e_{2} \wedge ... \wedge e_{n}.
\end{array}
\eq
Now it is possible to express the Pl\"ucker coordinates of the point in $Fl_n(\mathbb{R})$, which corresponds to a matrix $\Psi\in SO(n,\,\mathbb R)$:
\beq{point-fl}
y= (X_{i} \cdot e_{i}, \  X_{i_{1},i_{2}} \cdot e_{i_{1}} \wedge e_{i_{2}},..., \ X_{i_{1},i_{2},...,i_{m}} \cdot e_{i_{1}} \wedge e_{i_{2}} \wedge ... \wedge e_{i_{m}},...,  \ X_{i_{1},i_{2},...,i_{n-1}} \cdot e_{i_{1}} \wedge e_{i_{2}} \wedge ... \wedge e_{i_{n-1}}),
\eq
where $X_{i_{1},i_{2},...,i_{m}}$ are the Pl\"ucker coordinates expressible in the terms of the coordinates $\psi_{ij} \in \Psi$:
\beq{Plukk-1}
X_{i_{1},i_{2},...,i_{m}} = M_{\frac{1,2,...,m}{i_{1},i_{2},...,i_{m}}}(\psi),
\eq
where $M_{\frac{1,2,...,m}{i_{1},i_{2},...,i_{m}}}(\psi)$, as usual, denotes the minors obtained by the intersection of $m$ upper rows with some set of $m$ columns of the matrix $\Psi\in SO(n,\,\mathbb R)$.

We remind that it is possible to decompose the Lax operator $L \in Symm$ in the following form:
\beq{Decomp-GenToda-0}
L = \Psi \Lambda \Psi^{-1},
\eq
where $\Lambda$ is the diagonal eigenvalues matrix.

Note that
\beq{Plukk-2}
\begin{array}{c}
X^{\ast}_{i_{1},i_{2},...,i_{m}}= X_{j_{1},j_{2},...,j_{n-m}} = M_{\frac{1,2,...,n-m}{j_{1},j_{2},...,j_{n-m}}}(\psi) = (-1)^{a+b} \cdot M_{\frac{n-m+1,...,n}{i_{1},i_{2},...,i_{m}}}(\psi),\\
i_{k} \neq j_{l}, \  \ a=\sum 1+2+...+(n-m), \ b=\sum j_{1}+...+j_{n-m}
\end{array}
\eq
One says that Pl\"ucker coordinates belong to the same \textbf{\emph{family}}, if the corresponding minors of the matrix $\Psi\in SO(n,\,\mathbb R)$ have the same ranks. So $X$ and $X^{\ast}$ belong to the same family. It is possible to define the flag space using the quadratic relations on $X$ (see for example \cite{F} and Appendix A).
\beq{Quadratic-Relation}
\begin{array}{c}
X_{i_{1},i_{2},...,i_{m_{1}}} \cdot X_{j_{1},j_{2},...,j_{m_{2}}}-\sum X_{i^{'}_{1},i^{'}_{2},...,i^{'}_{m_{1}}} \cdot X_{j^{'}_{1},j^{'}_{2},...,j^{'}_{m_{2}}}=0,\\
\end{array}
\eq
where the sum is taking over the exchanging the first $k$ of the $j$ subscripts with $k$ of the $i$ subscripts, maintaining the order in each.

Note that except for the usual Pl\"ucker relations (the relations between the coordinates from the same family) there are additional quadratic relations  of $X$ which belong to the different families. No other relations of $X$ (except the relations (\ref{Quadratic-Relation})) exist.

It follows from (\ref{A-lambda-psi}) - (\ref{Integrals-Plukk}) that the integrals are rational functions of $\lambda$ and $\psi$. The numerators and the denominators of these functions consist of the sums such that each term of these sums is the product of some polynomial function of $\lambda$ and a function $\varphi(M(\psi))$. This function has the following form:
\beq{CrossRatio}
\varphi(M(\psi))=\frac{M_{\frac{1,2,...,m}{i_{1},i_{2},...,i_{m}}}M_{\frac{n-m+1,...,n}{i_{1},i_{2},...,i_{m}}}}{M_{\frac{1,2,...,m}{j_{1},j_{2},...,j_{m}}}M_{\frac{n-m+1,...,n}{j_{1},j_{2},...,j_{m}}}},
\eq

In the numerator and the denominator of $\varphi(M(\psi))$ the Pl\"ucker coordinate $X_{i_{1},i_{2},...,i_{m}}=M_{\frac{1,2,...,m}{i_{1},i_{2},...,i_{m}}}, \ X_{i_{1},i_{2},...,i_{m}} \in \mathbb{RP}^{C^{n}_{m}-1}$ is multiplied by the dual Pl\"ucker coordinate (from the same family) $X^{\ast}_{i_{1},i_{2},...,i_{m}}=M_{\frac{n-m+1,...,n}{i_{1},i_{2},...,i_{m}}}, \ X^{\ast}_{i_{1},i_{2},...,i_{m}} \in \mathbb({RP}^{C^{n}_{m}-1})^{\ast}$. We will call these products $\textbf{\emph{pairings}}$.

\begin{predl}
The functions $\ \varphi(M(\psi))$ (\ref{CrossRatio}) are integrals with respect to the one-parametric iso-spectral flows. The number of independent integrals in the set (\ref{CrossRatio}) is
\beq{FLdim}
N_{\Psi} = dim Fl_{n}(\mathbb{R}) - (n-1) \ .
\eq
\end{predl}

$\square$\\
The proof of this proposition is based on the direct calculation of the
$\varphi(M(\psi))$--dynamics using eqs. (\ref{minors-d}):
\beq{CrossRatio-3}
\begin{array}{c}
M^{'}_{\frac{1,2,...,m}{i_{1},i_{2},...,i_{m}}}=(-\sum_{l=1}^{m}a_{ll} + \sum_{i_{s}=i_{1}}^{i_{m}}\lambda_{i_{s}}) M_{\frac{1,2,...,m}{i_{1},i_{2},...,i_{m}}},\\
\ \\
M^{'}_{\frac{n-m+1,...,n}{i_{1},i_{2},...,i_{m}}}=(\sum_{l=n-m+1}^{n}a_{ll} - \sum_{i_{s}=i_{1}}^{i_{m}}\lambda_{i_{s}})M_{\frac{n-m+1,...,n}{i_{1},i_{2},...,i_{m}}} ,\\
\ \\
M^{'}_{\frac{1,2,...,m}{j_{1},j_{2},...,j_{m}}}=(-\sum_{l=1}^{m}a_{ll} + \sum_{j_{s}=j_{1}}^{j_{m}}\lambda_{j_{s}}) M_{\frac{1,2,...,m}{j_{1},j_{2},...,j_{m}}},\\
\ \\
M^{'}_{\frac{n-m+1,...,n}{j_{1},j_{2},...,j_{m}}}=(\sum_{l=n-m+1}^{n}a_{ll} - \sum_{j_{s}=j_{1}}^{j_{m}}\lambda_{j_{s}})M_{\frac{n-m+1,...,n}{j_{1},j_{2},...,j_{m}}}
\end{array}
\eq
and we get
\beq{CrossRatio-2}
\begin{array}{c}
\varphi^{'}(M(\psi))=0,\\
\end{array}
\eq
The same is true for the other flows (see (\ref{minors-d})).
The functions (\ref{CrossRatio}) arise from the relations between pairings of the different families of the Pl\"ucker coordinates, so these functions are defined on the flag space. As we showed above ((\ref{CrossRatio-2})-(\ref{CrossRatio-3})) these functions are invariants with respect to the flows induced by the iso-spectral integrals of motion $\frac{1}{k}TrL^{k}, k=\overline{2,n}$. The number of these flows is equal to $n-1$. It follows
from the invariant properties of the functions (\ref{CrossRatio}) that they are defined on the quotient space $FL_{n}(\mathbb{R})$ modulo the flow actions. The dimension of this quotient space is $dim Fl_{n}(\mathbb{R}) - (n-1)$ which is the number of independent non-homogeneous coordinates. So the number of the independent functions (\ref{CrossRatio}) is not greater than $dim Fl_{n}(\mathbb{R}) - (n-1)$. Note that in a general case the number of the functions(\ref{CrossRatio}) is greater or equal to $dim Fl_{n}(\mathbb{R}) - (n-1)$. On the other side the number of independent functions can not be less then $dim Fl_{n}(\mathbb{R}) - (n-1)$ because it is always possible to sort out the independent functions in such a way that they will $\textbf{not}$ satisfy the relation (\ref{Quadratic-Relation}). We know that there are no other relations on Pl\"ucker coordinates except (\ref{Quadratic-Relation}). Indeed, suppose that it is not true. Then we can choose $dim Fl_{n}(\mathbb{R}) - (n-1)$
ratios of the pairing but they are dependent. There exists some relation (at least one) on the ratios of the pairings. This relation is a equation on the independent Pl\"ucker coordinates and hence it is one of (\ref{Quadratic-Relation}) we get the contradiction. So we have
\beq{Intpsi}
\begin{array}{c}
N_{\psi}=n(n-1)/2-(n-1)=dim Fl_{n} - (n-1).
\end{array}
\eq
Note that we can choose the ratios in such a way that there is one independent non-homogeneous coordinate in each function (\ref{CrossRatio}). Why can we do it? First, the number $dim Fl_{n}(\mathbb{R}) - (n-1)$ is less (for $n \geq 4$) than the number of ratios of the pairings which is equal to $2^{n-1}-1-[\frac{n}{2}]$ and secondly each ratio of the pairings is the invariant and the product of two non-homogeneous coordinates. So each of $N_{\psi}$ ratios of the pairings depends on one non-homogeneous coordinate on the quotient space $FL_{n}(\mathbb{R})$ modulo the flow actions.
$\blacksquare$

Now in order to find the total number of the integrals of motion of the full symmetric Toda lattice it is necessary to subtract the number of Casimirs and to add the number of iso-spectral integrals of motion ($n-1$). In the even case the number of Casimirs is equal to $(n/2-1)$ and in the odd case -- $(n-1)/2$. So we get:
$$N = \frac{1}{2}n(n-1)- \frac{1}{2}n+1,$$
and
$$N = \frac{1}{2}n(n-1)- \frac{1}{2}(n+1) + 1$$
correspondingly for the even and odd cases. In general we have
\beq{formulas-fl-2}
\begin{array}{c}
N = \frac{1}{2}n(n-1)- [\frac{1}{2}(n+1)]+1.
\end{array}
\eq

The total number of the independent integrals of motion of the form (\ref{A-lambda-psi-Integrals}) is determined by the number of independent integrals of motion. Each of these integrals is obtained from the ratio of semi-invariants (\ref{A-lambda-psi}) and some chosen paring for each family.
\beq{A-lambda-psi-2}
\frac{A^{(k)}_{\frac{n-m+1,...,n}{1,2,...,m}}}{M_{\frac{1,2,...,m}{a_{1},a_{2},...,a_{m}}}M_{\frac{n-m+1,...,n}{a_{1},a_{2},...,a_{m}}}}=\sum_{i_{1},i_{2},...,i_{m}} \lambda^{k}_{i_{1}}\lambda^{k}_{i_{2}} \cdot...\cdot \lambda^{k}_{i_{m}} \frac{M_{\frac{1,2,...,m}{i_{1},i_{2},...,i_{m}}}M_{\frac{n-m+1,...,n}{i_{1},i_{2},...,i_{m}}}}
{M_{\frac{1,2,...,m}{a_{1},a_{2},...,a_{m}}}M_{\frac{n-m+1,...,n}{a_{1},a_{2},...,a_{m}}}},\\
\eq
where we chose some $i_{l}=a_{l}$ and fix these indices.
The number of these invariants in its turn is determined by the number of the independent invariants (\ref{CrossRatio}). The invariants linearly (\ref{A-lambda-psi-2}) depend on the invariants (\ref{CrossRatio}).
So the total number of the independent integrals of motion expressed by the matrix elements of the Lax and having the form (\ref{Integrals-Plukk}) -- (\ref{A-lambda-psi-Integrals}) is defined by the formula (\ref{formulas-fl-2}) and we have found explicit expressions for the integrals of motion of the full symmetric Toda lattice.

Now let us define how one can choose a non-commutative set of the independent integrals of motion. First of all let us give the answer to the question: how many independent integrals of motion we can choose from the integrals constructed from the pairings which belong to the same family? The Pl\"ucker coordinates from the same family describe the partial flag space
\beq{pfl}
\begin{array}{c}
Q_{pfl}(\mathbb{R})= V^{a} \subset V^{b} \subset V^{n}= \mathbb{R}^{n},\\
Q_{pfl}(\mathbb{R}) \hookrightarrow \mathbb{RP}^{C^{n}_{m}-1} \times (\mathbb{RP}^{C^{n}_{n-m}-1})^{\ast},\\
a = m, \  \mbox{dim}V^{a} = m,\\
b = n - m, \  \mbox{dim}V^{b} = n - m.
\end{array}
\eq
Note that severally each embedding $V^{a} \subset V^{n}= \mathbb{R}^{n}$ and $V^{b} \subset V^{n}= \mathbb{R}^{n}$ give Grassmanians $Gr_{m}(n)$ and $Gr_{n-m}(n)$. In order to calculate the number of the independent variables of the partial flag space $Q_{pfl}(\mathbb{R})$ it is necessary to calculate the co-dimension of stabilizer $B_{pfl}$ because $Q_{pfl}(\mathbb{R}) \cong G/B_{pfl}$ where $G = GL_{n}(\mathbb{R})$ (or $G = SO_{n}(\mathbb{R})$). Also we have to subtract the number of flows induced by the iso-spectral integrals of motion. So we get
\beq{NpartFl}
\begin{array}{c}
N_{pfl}=2m(n-m)-m^{2}-(n-1)=,\\
=n(2m-1)-3m^{2}+1,\\
m \leq [ \frac{n}{2} ].
\end{array}
\eq
\begin{figure}[!t]
\begin{center}
\includegraphics[width=250pt,height=270pt]{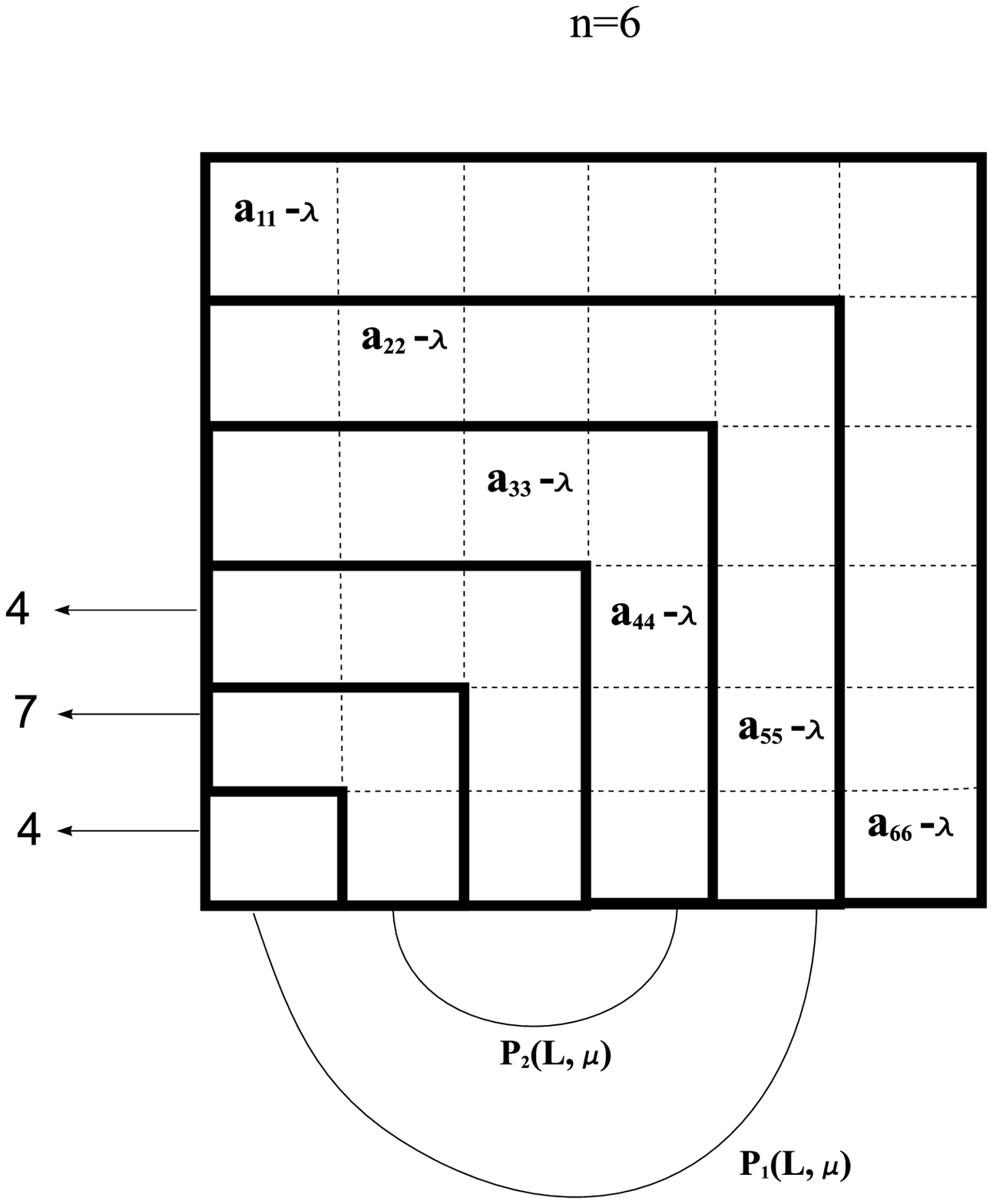}
\end{center}
\end{figure}
\paragraph{For example let us consider the case $n=6$}. There are the following families:\\
-- the first family is formed by the Pl\"ucker coordinates which are the determinants of the submatrix of sizes $1\times1$ and $5\times5$,\\
-- the second family is formed by the Pl\"ucker coordinates which are the determinants of the submatrix of sizes $2\times2$ and $4\times4$,\\
-- the third family is formed by the Pl\"ucker coordinates which are the determinants of the submatrix of size $3\times3$
(formula (\ref{NpartFl}) is correct in this case too). Let us find estimates for the number of integrals.
\beq{Restr}
\begin{array}{c}
1: N^{1}_{pfl}=6(2-1)-3+1=4,\\
2: N^{2}_{pfl}=6(2\cdot2-1)-3\cdot4+1 = 18-12+1=7,\\
3: N^{3}_{pfl}=6(6-1)-3\cdot9+1=30-27+1=4,\\
\end{array}
\eq
Note that these estimates account for the number of Casimirs ($2$ in the case $n=6$).\\

So, to form a non-commutative set of independent integrals (maximal number of the invariants is equal to $16$ including $3$ Casimirs and
$5$ iso-spectral integrals, the formula for the maximal number of invariants is $N_{max}=\frac{1}{2}n(n-1)+1$) we need to choose $8$ functions of the form (\ref{A-lambda-psi-Integrals}) from $4+7+4=15$ functions so that there are no more than $4$ functions corresponding
to submatrix $A$ of the size $1 \times 1$, no more than $7$ functions corresponding to submatrix $A$ of the size $2 \times $ and no more than $4$
functions corresponding to submatrix $A$ of the size $3 \times 3$. Choosing $8$ functions we must take into account that if we put the index $k_{2}=1$ the index $k_{1}$ must be more or equal to $3$ in order to exclude the Casimirs from the non-commutative set. Adding $5$ functions of the form $\frac{1}{k}TrL^{k}, \ k \geq 2$ we get the required non-commutative set. In general case we must act similarly to choose the integrals for the non-commutative set of independent integrals in terms of matrix elements of the Lax.

We have schematically described at the picture $n=6$ the maximal number of the integrals in the case $n=6$. The arrows at the left side point at the maximal number of the integrals for each family. They go from the top elements of the matrices which form the integrals and semi-invariants of motion in the corresponding family. The lines below denote the spectral curves which obtained by chopping procedure. Note that all statements concerning the non-commutative set of the independent integrals in the case $n=6$ were checked by direct computations with the help of Mathematica 8.

\paragraph{Full non-involutive set of independent integrals}
Now we are ready to describe explicitly the full non-involutive set of independent integrals expressed in terms of matrix elements of the Lax operator $L$. For this goal one can choose the following integrals from the set (\ref{A-lambda-psi-Integrals}):
\beq{A-lambda-psi-Integrals-2}
\begin{array}{c}
J_{k,1}= \frac{A^{(k)}_{\frac{n-m+1,...,n}{1,2,...,m}}}{A_{\frac{n-m+1,...,n}{1,2,...,m}}} \ \ \ \mbox{with}\\
m=1\ , \  \ k=\overline{3,n-1} \  \ \ \ \ \ \ \ \ \ \ \ \ (\# \ \mbox{integrals} = n-3) \\
 2 \leq m < [\frac{n}{2}]\ , \ k=\overline{2,n-2m} \ \ \ (\# \ \mbox{integrals} = [\frac{(n-4)^{2}}{4}])\\
 m=[\frac{n}{2}]\ , \ \ k=\overline{2,[\frac{(n-2)^2}{4}]+1} \ \ \ \ \ (\# \ \mbox{integrals} = [\frac{(n-2)^2}{4}])
\end{array}
\eq
together with the $(n-1)$ iso-spectral integrals $\frac{1}{k}TrL^{k}, k=\overline{2,n}$, so that the total number of the integrals is given by eq. (\ref{Integrals-number}).

In order to form the full non-involutive set of independent integrals expressed in terms of the eigenvector matrix $\Psi$, one should for every $m$ from the range $1 \leq m < [\frac{n}{2}]$ select $n-2m-1$ integrals from the set (\ref{CrossRatio}) and for $m=[\frac{n}{2}]$ select $[\frac{(n-2)^2}{4}]$ integrals from the set (\ref{CrossRatio}). The union of these integrals together with the $n-1$ integrals expressed in terms of the eigenvalue matrix $\Lambda$ gives a full non-involutive set of independent integrals of the full symmetric $\mathfrak{sl}_n$ Toda lattice, which number is $N_{n}$ (\ref{Integrals-number}).

\subsection{Families of the integrals in involution in the case $n=5$}

Let us consider the case $n=5$ in more detail and obtain the expressions for the integrals of motion in an explicit form. The matrix of the Lax operator has the following form:
\beq{L5-1} L=\left(
\begin{array}{c c c c c}
 a_{11} & a_{12} & a_{13} & a_{14} & a_{15}\\
 a_{21} & a_{22} & a_{23} & a_{24} & a_{25}\\
 a_{31} & a_{32} & a_{33} & a_{34} & a_{35}\\
 a_{41} & a_{42} & a_{43} & a_{44} & a_{45}\\
 a_{51} & a_{52} & a_{53} & a_{54} & a_{55}\\
\end{array}
\right),
\eq
From the chopping procedure we have the following spectral curves:
\beq{Pk5-2}
\begin{array}{c}
P_{0}(L,\mu) = det(L-\mu I),\\
P_{1}(L,\mu) = det(L-\mu I)_{1},\\
P_{2}(L,\mu) = det(L-\mu I)_{2},\\
\end{array}
\eq

\beq{Pk5}
\begin{array}{c}
P_{0}(L,\mu) = -\mu^{5} + \mu^{4}TrL - \mu^{3}(\frac{1}{2}(TrL)^{2} - \frac{1}{2}TrL^{2}) + \mu^{2} (\frac{1}{3}(TrL)^{3} - \frac{1}{3}TrL^{3})-\\
- \mu (\frac{1}{4}(TrL)^{4} - \frac{1}{4}TrL^{4}) + detL,\\
\ \\
P_{1}(L,\mu) = \mu^{3}a_{15} + \mu^{2} (A_{\frac{25}{12}} + A_{\frac{45}{14}} + A_{\frac{35}{13}}) + \mu (A_{\frac{345}{134}} + A_{\frac{245}{124}} + A_{\frac{235}{123}}) + A_{\frac{2345}{1234}},\\
\ \\
P_{2}(L,\mu) = -\mu A_{\frac{45}{12}} + A_{\frac{345}{123}}.
\end{array}
\eq

The picture $n=5$ below describes the spectral curves obtained by chopping procedure.
\begin{figure}[!t]
\begin{center}
\includegraphics[width=250pt,height=270pt]{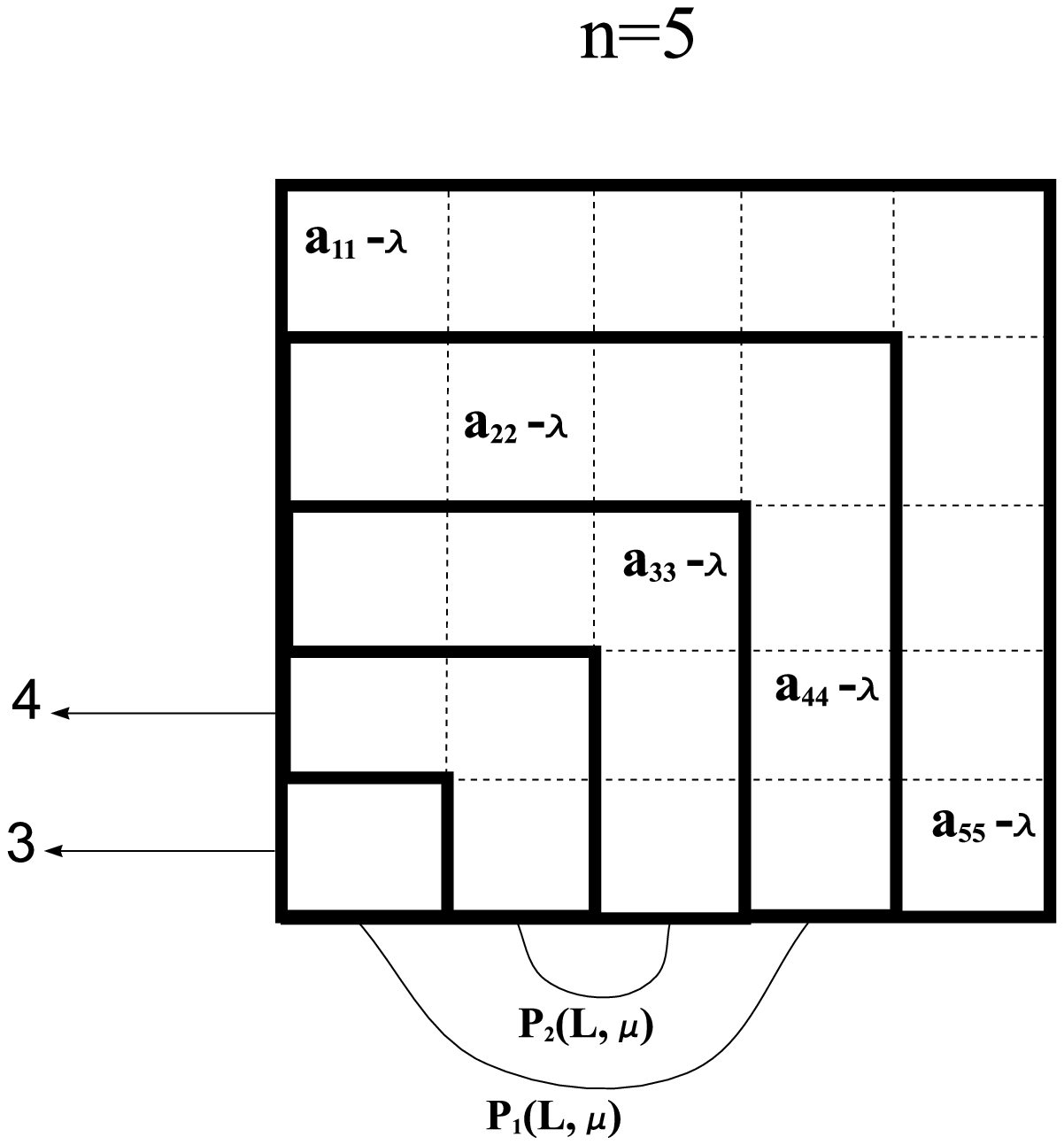}
\end{center}
\end{figure}
The upper arrow at the left side points at the maximal number of the integrals constructed using the semi-invariant $A_{\frac{45}{12}}$. These integrals take the following form
\beq{J-5}
\begin{array}{c}
J_{1}= \frac{A^{(2)}_{\frac{45}{12}}}{A_{\frac{45}{12}}}=\frac{\sum_{i}\sum_{j} \lambda^{2}_{i}\lambda^{2}_{j} (-1)^{i+j+1} M_{\frac{12}{ij}}M_{\frac{123}{klm}}}{\sum_{i}\sum_{j} \lambda_{i}\lambda_{j} (-1)^{i+j+1} M_{\frac{12}{ij}}M_{\frac{123}{klm}}},  \\

\ \\

J_{2}= \frac{A^{(3)}_{\frac{45}{12}}}{A_{\frac{45}{12}}}=\frac{ \sum_{i}\sum_{j} \lambda^{3}_{i}\lambda^{3}_{j} (-1)^{i+j+1} M_{\frac{12}{ij}}M_{\frac{123}{klm}}}{ \sum_{i}\sum_{j} \lambda_{i}\lambda_{j} (-1)^{i+j+1} M_{\frac{12}{ij}}M_{\frac{123}{klm}}},\\

\ \\

\ i<j, \ \ \ i,j =\overline{1,5}.
\end{array}
\eq
The lower arrow at the left side points at the maximal number of the integrals constructed using the semi-invariant $a_{15}$ and the corresponding integrals have the following form:
$$ I_{1,1} = \frac{A_{\frac{25}{12}} + A_{\frac{45}{14}} + A_{\frac{35}{13}}}{a_{15}}, \ \ \ \ \ I_{2,1} = \frac{ A_{\frac{345}{134}} + A_{\frac{245}{124}} + A_{\frac{235}{123}}}{a_{15}} ,  \ \ \ \ \ I_{3,1} = \frac{A_{\frac{2345}{1234}}}{a_{15}}.$$

$$ I_{1,2} = \frac{A_{\frac{345}{123}}}{A_{\frac{45}{12}}}.$$

The first family in involution consist of the following integrals of motion
\beq{family-1}
\begin{array}{c}
\frac{1}{2}Tr L^{2}, \ \frac{1}{3}Tr L^{3}, \ \frac{1}{4}Tr L^{4}, \ \frac{1}{5}Tr L^{5}, \ I_{2,1}, \ I_{3,1}.
\end{array}
\eq
The second family in involution consist of the following integrals of motion
\beq{family-1}
\begin{array}{c}
\frac{1}{2}Tr L^{2}, \ \frac{1}{3}Tr L^{3}, \ \frac{1}{4}Tr L^{4}, \ \frac{1}{5}Tr L^{5}, \ J_{1}, \ J_{2}.
\end{array}
\eq
The number of Casimirs is equal to $3$: $I_{1,1}, \ I_{1,2}$ and $TrL$. The total number of the integrals (non-commutative family) is equal to $8$:
\beq{formulas-5}
\begin{array}{c}
N = \frac{1}{2}n(n-1)- [\frac{1}{2}(n+1)]+1,  \\
N_{5} = \frac{1}{2}5(5-1)- [\frac{1}{2}(5+1)]+1 = 10-3+1 = 8.
\end{array}
\eq
Note that instead of the integrals $I_{2,1}, \ I_{3,1}$ we can consider the following integrals:
$$I^{51}_{1} = \frac{A^{3}_{51}}{a_{15}}=\frac{ \sum_{i} \lambda^{3}_{i}\psi_{1i}\psi_{5i} }{ \sum_{i} \lambda_{i}\psi_{1i}\psi_{5i}},$$

$$I^{51}_{2} = \frac{A^{4}_{51}}{a_{15}}=\frac{ \sum_{i} \lambda^{4}_{i}\psi_{1i}\psi_{5i} }{ \sum_{i} \lambda_{i}\psi_{1i}\psi_{5i}},$$
and also the Casimir
$$I^{51}_{0} = \frac{A^{2}_{51}}{a_{15}}=\frac{ \sum_{i} \lambda^{2}_{i}\psi_{1i}\psi_{5i} }{ \sum_{i} \lambda_{i}\psi_{1i}\psi_{5i}}, \ \ \ I_{1,1}=I^{51}_{0} - TrL.$$

So we can say that the first family in involution is induced by the minor $a_{15}$ and the second one by the minor $A_{\frac{45}{12}}$.
It is analogous to the case $n=4$ ($a_{14}$ and $A_{\frac{34}{12}}$ correspondingly).

At the end of this section we bring forward the table for the integrals of the full symmetric Toda lattice for different $n$:
\beq{t2}
\scriptsize{
\begin{tabular}{|c|c|c|c|c|}
\hline\cline{1-0}
$$ & $$ & $$ & $$ & $$\\
$\ \ \ n \ \ \ $ & $\ \ \ dim(Fl_{n}) \ \ \ $ & $ \ \ \ N_{n}^{Iso} \ \ \ $ & $ \ \ \ N_{n}^{Chopp}+N_{n}^{Add} \ \ \ $ & $ \ \ \ N_{n} \ \ \ $\\

$$ & $$ & $$ & $$ & $$\\
\hline\cline{1-0}
$$ & $$ & $$ & $$ & $$\\
$4$ & $6$ & $3$ & $1+1$ & $5$\\

$$ & $$ & $$ & $$ & $$\\
\hline\cline{1-0}
$$ & $$ & $$ & $$ & $$\\
$5$ & $10$ & $4$ & $2+2$ & $8$\\

$$ & $$ & $$ & $$ & $$\\
\hline\cline{1-0}
$$ & $$ & $$ & $$ & $$\\
$6$ & $15$ & $5$ & $4+4$ & $13$\\

$$ & $$ & $$ & $$ & $$\\
\hline\cline{1-0}
$$ & $$ & $$ & $$ & $$\\
$7$ & $21$ & $6$ & $6+6$ & $18$\\

$$ & $$ & $$ & $$ & $$\\
\hline\cline{1-0}
$$ & $$ & $$ & $$ & $$\\
$8$ & $28$ & $7$ & $9+9$ & $25$\\

$$ & $$ & $$ & $$ & $$\\
\hline
\end{tabular}
\ ,
}
\eq
where\\
$n$ -- the order of the Lax,\\
$dim(Fl_{n})$ is the dimension of the flag space,\\
$N_{n}^{Iso}$ is the number of the iso-spectral integrals of motion minus Casimir,\\
$N_{n}^{Chopp}+N_{n}^{Add}$ is the number of the invariants of motion (minus Casimirs) obtained by chopping procedure and additional integrals,\\
$N_{n}$ is the total number of the integrals in non-commutative family.\\

\paragraph{Conclusions}
In this paper we developed a new approach to derive integrals of motion of the full symmetric $\mathfrak{sl}_n$ Toda lattice which uncovers its "genetics" from the viewpoint of flag spaces. We use the semi-invariants (\ref{minors-d}), which are Pl\"ucker coordinates (\ref{Plukk-1}) in the corresponding projective spaces, in order to construct explicitly the full set of the non-involutive integrals expressed both in terms of the Lax matrix (\ref{Lax}) and its eigenvalue and eigenvector matrices (\ref{CrossRatio}) of arbitrary ranks.

Our approach is much simpler than the one based on Kostant procedure \cite{BG-1,GS,BG-2} and avoids the crucial computational complexities
appearing in the latter procedure even for low-rank Lax matrices, which prevent it use for the higher ranks.

The simplicity of the advocated approach is exemplified by the additional integral $J={A^{(2)}_{\frac{3,4}{1,2}}}(A_{\frac{3,4}{1,2}})^{-1}$ of the full $\mathfrak{sl}_4$ Kostant-Toda lattice. In order to derive it the authors of \cite{EFS,S2} applied the isomorphism $\mathfrak{sl}_4
\leftrightarrow \mathfrak{so}_6$ and the $\mathfrak{so}_6$-chopping procedure.

The results of the present paper are crucial to establish the Bruhat order in the full symmetric $\mathfrak{sl}_n$ Toda lattice \cite{CSS}. The short version of our paper is represented in \cite{CS}. The technique that we have developed in the present paper has further extensions and applications. The generalization to the Toda lattices defined for other Lie algebras and homogeneous spaces will be given elsewhere.

\paragraph{Acknowledgments}
The authors would like to thank G.I. Sharygin and D. Sternheimer for the fruitful discussions and remarks.
The work of Yu.B. Chernyakov was supported in part by grants RFBR-12-02-00594 and by the Federal Agency for Science and Innovations of Russian Federation under contract 14.740.11.0347. The work of A.S. Sorin was supported in part by the RFBR Grants No. 11-02-01335-a, No. 13-02-91330-NNIO-а and No. 13-02-90602-Arm-a.

\section{Appendix}

\subsection{Appendix A. Description of the flag space}
Let consider the subgroup of the upper triangular matrices $GL(n, \mathbb{R})$ in the case $n=4$:
\beq{Stab-x-Fl} B^{+}= \left(
\begin{array}{c c c c}
  b_{11} & b_{12} & b_{13} & b_{14}\\
 0 & b_{22} & b_{23} & b_{24}\\
 0 & 0 & b_{33} & b_{34}\\
 0 & 0 & 0 & b_{44}\\
\end{array}
\right),
 \eq
It turns out that the quotient $GL(4,\mathbb{R}) / B^{+}$ is isomorphic to the flag space.
Recall that the points of the flag space (flags) are the sequences of the linear embeddings:$V^{0} \hookrightarrow V^{1} \hookrightarrow V^{2} \hookrightarrow V^{3} \hookrightarrow \mathbb{R}^{4}$.
Embedding each subspace of the flag in some projective space it is possible to show that flag space embeds in the direct products of the projective spaces:
\beq{map-fl}
\begin{array}{c}
GL(4,\mathbb{R}) / B^{+} \hookrightarrow \mathbb{RP}^{3} \times \mathbb{RP}^{5} \times (\mathbb{RP}^{3})^{\ast},\\
e_{i} \wedge e^{\ast}_{j} = e_{1} \wedge e_{2} \wedge e_{3} \wedge e_{4}.
\end{array}
\eq
It gives us the possibility to describe the points of flag space with the help of projective coordinates.
In order to construct the map $Fl_4\leftrightarrows GL(4, \mathbb{R})/B^+$ let us choose the point $x= (e_{1}, e_{1} \wedge e_{2}, e_{1} \wedge e_{2} \wedge e_{3})$ of the flag space
(by the external product of the vectors we denote the space spanned on these vectors).
In terms of the matrices this point corresponds to the identity matrix of order $4\times4$.
The stabilizer of this point is the subgroup $B^+$.
To see this, one can compare the projective coordinates of the point $x$ and its image under the subgroup action:
\beq{transf-fl}
\begin{array}{c}
v_{1}=B_{x} e_{1}=b_{11}e_{1},\\
v_{2}=B_{x} e_{2}=b_{12}e_{1}+b_{22}e_{2},\\
v_{3}=B_{x} e_{3}=b_{13}e_{1}+b_{23}e_{2}+b_{33}e_{3},\\
x^{'}=( b_{11} \cdot e_{1}, \ b_{11}b_{22} \cdot e_{1} \wedge e_{2}, \ b_{11}b_{22}b_{33} \cdot e_{1} \wedge e_{2} \wedge e_{3}).
\end{array}
\eq
So the point $x^{'}$ defines the same flag as the point $x$ and the isomorphism $GL(4,\, \mathbb{R})/B^+\cong Fl_4$
follows from the general theory of homogeneous spaces.
Now it is possible to express the projective coordinates of a point in a general position in $Fl_4$ corresponding to the matrix
$\Psi \in GL(4,\, \mathbb{R})$:
\beq{flag-psi} \Psi = \left(
\begin{array}{c c c c}
 \psi_{11} & \psi_{12} & \psi_{13} & \psi_{14}\\
 \psi_{21} & \psi_{22} & \psi_{23} & \psi_{24}\\
 \psi_{31} & \psi_{32} & \psi_{33} & \psi_{34}\\
 \psi_{41} & \psi_{42} & \psi_{43} & \psi_{44}\\
\end{array}
\right),
 \eq
namely:
\beq{genpos-fl}
y= (X_{i} \cdot e_{i}, \ X_{ij} \cdot e_{i} \wedge e_{j}, \ X_{ijk} \cdot e_{i} \wedge e_{j} \wedge e_{k}),
\eq
where
\beq{Plukk-A}
X_{i_{1},i_{2},...,i_{m}} = M_{\frac{1,2,...,m}{i_{1},i_{2},...,i_{m}}}(\psi).
\eq
The dimension $Fl_4$ can be calculated from the isomorphism $GL(4,\, \mathbb{R})/B^+\cong Fl_4$, it is equal to $\frac{n(n-1)}{2}$.
 The flag space is a variety points of which satisfy a system of quadratic equations in the product of the projective spaces. In our case we have
\beq{map-fl-2}
\begin{array}{c}
GL(4,\mathbb{R}) / B^{+} \hookrightarrow W=\mathbb{RP}^{3} \times \mathbb{RP}^{5} \times (\mathbb{RP}^{3})^{\ast},
\end{array}
\eq
where $W$ has the dimension $11$, and dimension of $Fl_4$ must be equal to $6$. The system of Pl\"ucker equations has the following form:
\beq{Syst-eq-Plukk}
\begin{array}{c}
X_{1} \cdot X_{234} - X_{2} \cdot X_{134} + X_{3} \cdot X_{124} - X_{4} \cdot X_{123} = 0,\\
\ \\
X_{1} \cdot X_{23} - X_{2} \cdot X_{13} + X_{3} \cdot X_{12} = 0,\\
X_{1} \cdot X_{24} - X_{2} \cdot X_{14} + X_{4} \cdot X_{12} = 0,\\
X_{1} \cdot X_{34} - X_{3} \cdot X_{14} + X_{4} \cdot X_{13} = 0,\\
X_{2} \cdot X_{34} - X_{3} \cdot X_{24} + X_{4} \cdot X_{23} = 0,\\
\ \\
X_{12} \cdot X_{34} - X_{13} \cdot X_{24} + X_{14} \cdot X_{23} =0,\\
\ \\
X_{13} \cdot X_{234} - X_{23} \cdot X_{134} + X_{34} \cdot X_{123} = 0,\\
X_{12} \cdot X_{234} - X_{23} \cdot X_{124} + X_{24} \cdot X_{123} = 0,\\
X_{14} \cdot X_{234} - X_{24} \cdot X_{134} + X_{34} \cdot X_{124} = 0,\\
X_{14} \cdot X_{123} - X_{13} \cdot X_{124} + X_{12} \cdot X_{134} = 0.\\
\end{array}
\eq
\ \\
This system has $5$ independent equation. The number of Pl\"ucker coordinates is equal to $14$.
Passing to the non-homogeneous coordinates we must subtract $3$ (one for every projective space). So we have
$$dimFl_4=14-3-5=6,$$
just as has to be. In the general case the number of the independent equation is given by the following formula:
$$
N_{eq} = 2^{n} - 1 - \frac{n(n+1)}{2}.
$$
In the general case the number of the Pl\"ucker coordinates is equal to $2^{n}-2$.

The same reasoning can be used when the flag space is identified with the quotient $SO(n, \mathbb{R})/S$,
 where the subgroup $S$ of the group $SO(n, \mathbb{R})$ consists of the diagonal matrices with the diagonal elements $\pm 1$ and $det S = 1$.

It follows from (\ref{Intpsi}) that the number of independent integrals is equal to $N_{\psi}=dim Fl_{n} - (n-1)=6-3=3$. The number of ratios of the pairings is equal to $5$, for example, $\frac{X_{1} \cdot X_{234}}{X_{4} \cdot X_{123}}, \ \frac{X_{2} \cdot X_{134}}{X_{4} \cdot X_{123}}, \ \frac{X_{3} \cdot X_{124}}{X_{4} \cdot X_{123}}, \ \frac{X_{12} \cdot X_{34}}{X_{14} \cdot X_{23}}, \ \frac{X_{13} \cdot X_{24}}{X_{14} \cdot X_{23}}$. From (\ref{Syst-eq-Plukk}) we conclude that the number of equation for the ratios of the pairings is equal to $2$. So we can chose the independent invariants as needed.

\subsection{Appendix B. Dynamics on the flag space}

In this subsection we follow the papers \cite{EFS} and \cite{S2}
to show why in the case of full Kostant-Toda lattice the action of the one-parametric subgroup induced by the iso-spectral integrals of motion is diagonalized on the flag spaces.
Also we consider the case of the full symmetric Toda lattice and show the distinction between the dynamics on the flag space of these two systems.

\paragraph{Full Kostant-Toda lattice}

Let us consider the full Kostant-Toda lattice. The Lax has the following form:
\beq{LaxfCT} X=\left(
\begin{array}{c c c c c}
 x_{11} & 1 & 0 & ... & 0\\
 x_{21} & x_{22} & 1 & ... & 0\\
 ... & ... & ... & ... & ...\\
  x_{n-1 \ 1} & ... & ... &  x_{n-1 \ n-1} & 1\\
 x_{n1} &  ... & ... & x_{n \ n-1} & x_{nn}
 \end{array}
\right), \ \ \ X = \epsilon + \beta_{-}, \ \ \ X \in \mathfrak{sl}_n(\mathbb{C}).
\eq
It turns out that that these matrices can be expressed as the following product (\cite{K}):
$$X = LCL^{-1},$$
where $L$ is the unique unipotent matrix for each $X$,
\beq{matrixC}
C=\left(
\begin{array}{c c c c c}
 0 & 1 & 0 & ... & 0\\
 0 & 0 & 1 & ... & 0\\
 ... & ... & ... & ... & ...\\
 0 & ... & ... &  0 & 1\\
 s_{n} &  ... & ... & s_{2} & 0
 \end{array}
\
\right),
\eq
and $s_{i}$ are the coefficients of characteristic polynomial of the matrix $X$:
$$\lambda^{n} - s_{2}\lambda^{n-2} - ... - s_{n} = 0.$$
In the generic case (all $\lambda$ are different) as the result of expansion we get:
\beq{Decomp-fCT}
X = LV \Lambda V^{-1}L^{-1}, \ Z = LV,
\eq
where $V$ is the Vandermonde matrix, $\Lambda$ is the eigenvalue matrix.
Note that $X$ and $Z$ belong to $\mathfrak{sl}_n(\mathbb{C})$. Each matrix $Z^{-1}$ defines a flag.
Fixing eigenvalues $\lambda$ we get an embedding of the orbit of element $X$ in the flag space
\beq{map-fCT}
X \mapsto Z^{-1} mod B_{+}.
\eq
Let us calculate the dynamics on the flag space induced by the integrals $\frac{1}{k}TrX^{k}, \ k \geq 2$.
\beq{DinfCTflag-1}
\begin{array}{c}
X(t_{k}) = Ad^{\ast}_{n_{-}(t_{k})}X(0) = n^{-1}_{-}(t_{k}) X(0) n_{-}(t_{k}), \ \ \ \exp(t_{k} \nabla \frac{1}{k}TrX^{k}(0)) = n_{-}(t_{k})b_{+}(t_{k}),\\
\ \\
\nabla \frac{1}{k}TrX^{k}(0) = X^{k-1}(0), \\
\ \\
X(t_{k}) = Z(t_{k}) \Lambda Z^{-1}(t_{k}), \ \ \ Z^{-1}(t_{k}) = V^{-1}L^{-1} n_{-}(t_{k}) = Z^{-1}(0)n_{-}(t_{k}).
\end{array}
\eq
Here we use the factorization theorem (see, for example, \cite{FT}) which allows one to express the solution of the equation of motion
\beq{eq-of-motion-fCT}
X^{'}(t) = [X(t), \sqcap_{\mathcal{N_{-}}} X(t)], \ \ \ sl(n,\mathbb{C}) = \beta_{+} + \mathcal{N_{-}},
\eq
in the form
\beq{eq-of-motion-fCT}
X(t) = Ad^{\ast}_{n_{-}(t)}X(0).
\eq

As the result we get
\beq{DinfCTflag-2}
\begin{array}{c}
Z^{-1}(t_{k}) mod B_{+} =\\
\ \\
= V^{-1}L^{-1} n_{-}(t_{k}) mod B_{+} = V^{-1}L^{-1} n_{-}(t_{k})b_{+}(t_{k}) mod B_{+} =\\
\ \\
=  V^{-1}L^{-1} \exp(t_{k} X^{k-1}(0))  mod B_{+} = V^{-1}L^{-1} \exp(t_{k} (LV \Lambda V^{-1}L^{-1})^{k-1})  mod B_{+} =\\
\ \\
= V^{-1}L^{-1} \exp(LV t_{k} \Lambda^{k-1} V^{-1}L^{-1})  mod B_{+} =  V^{-1}L^{-1} LV \exp(t_{k} \Lambda^{k-1}) V^{-1}L^{-1}  mod B_{+} = \\
\ \\
= \exp(t_{k} \Lambda^{k-1}) V^{-1}L^{-1}  mod B_{+} = \exp(t_{k} \Lambda^{k-1}) Z^{-1}(0) mod B_{+}.
\end{array}
\eq
where we use the equality $\exp(B \cdot A \cdot B^{-1}) = B \cdot \exp A \cdot B^{-1}$.
The flows $X(t_{k})$ induce the action of the toric group $(\mathbb{C}^{\ast})^{n-1}.$

\paragraph{Full symmetric Toda lattice}
Let us show that we cannot define the embedding $L \in Symm_{n}$ in the flag space so that the induced dynamics on the flag can be diagonalized.
\beq{Decomp-GenToda}
L = \Psi \Lambda \Psi^{-1}, \ \Psi \in SO(n, \mathbb{R}).
\eq
The matrix of the Lax is defined up to the action of diagonal subgroup $S \in SO(n, \mathbb{R})$ on $\Psi$.
The flag space is $SO(n, \mathbb{R})/S$. Each matrix $\Psi^{-1}$ defines a flag.
Let $S$ act on $\Psi^{-1}$ from the left.
It follows from the invariance of the Lax under the action $\Psi \rightarrow \Psi S$ that this action changes the flag space because
it permutes the classes of $SO(n, \mathbb{R})/S$ and we cannot uniquely associate a flag to the Lax:
\beq{Flag-GenToda}
\begin{array}{c}
L = \Psi S \Lambda S^{-1} \Psi^{-1}\\
\ \\
S_{1}^{-1} \Psi_{a}^{-1} mod S \neq S_{2}^{-1} \Psi_{a}^{-1} mod S.
\end{array}
\eq
We see that several different flags can be associated with one matrix of the Lax.


\begin{thebibliography}{60}

\bibitem{T1}
M. Toda, Vibration of a chain with nonlinear interaction, J. Phys. Soc. Japan 22(2) (1967), 431 -- 436.

\bibitem{T2}
M. Toda, Wave propagation in anharmonic lattices, J. Phys. Soc. Japan 23(3) (1967), 501 -- 506.

\bibitem{H}
M. Henon, Integrals of the Toda lattice, Phys. Rev. B9 (1974), 1921 -- 1923.

\bibitem{F1}
H. Flaschka, The Toda lattice. I. Existence of integrals, Phys. Rev. B 9(4) (1974), 1924 -- 1925.

\bibitem{F2}
H. Flaschka, On the Toda lattice. II. Prog. Theor. Phys. 51(3) (1974), 703 -- 716.

\bibitem{BG-1}
A. M. Bloch and M. Gekhtman, Hamiltonian and gradient structures in the Toda flows, J. Geom. Phys. 27 (1998), 230 -- 248.

\bibitem{Neh}
N.N. Nehoroshev, Action-angle variables and their generalization,  Tr. Mosk. Mat. O.-va. 26 (1972), 181 -- 198.

\bibitem{GS}
M. Gekhtman and M. Shapiro, Noncommutative and commutative integrability of generic Toda
flows in simple Lie algebras, Comm. Pure and Appl. Math. 52  (1999), 53 -- 84.

\bibitem{BG-2}
A. M. Bloch and M. Gekhtman, Lie algebraic aspects of the finite nonperiodic Toda
flows, J. Comp. Appl. Math. 202 (2007), 3 -- 25.

\bibitem{A}
A.A. Arhangelskii, Completely integrable hamiltonian systems on a group of triangular matrices, Mathematics of the USSR-Sbornik (1980), 36:1, 127 -- 134.

\bibitem{Ad}
M. Adler, On a trace functional for pseudo-differential operators and the symplectic structure of the Korteweg-de Vries equation, Invent. Math., 50 (1979), 219 -- 248.

\bibitem{K1}
B. Kostant, The solution to a generalized Toda lattice and representation theory, Adv. in Math. 34 (1979), 195 -- 338.

\bibitem{S}
W. W. Symes, Systems of Toda type, inverse spectral problems, and representation theory, Invent. Math. 59 (1980), no. 1, 13 -- 51.

\bibitem{DLNT}
P. Deift, L. C. Li, T. Nanda, and C. Tomei, The Toda flow on a generic orbit is integrable, CPAM 39 (1986), 183 -- 232.

\bibitem{EFS}
N. Ercolani, H. Flaschka, and S. Singer, The geometry of the full Kostant-Toda lattice In: Integrable Systems,
Vol. 115 of Progress in Mathematics, Birkhauser (1993), 181 -- 226.

\bibitem{FS}
P. Fre, A.S. Sorin, The arrow of time and the Weyl group: all supergravity billiards are integrable, Nucl. Phys., B 815 (2009), 430,[arXiv:0710.1059].

\bibitem{S1}
B. A. Shipman, On the geometry of certain isospectral sets in the full Kostant-Toda lattice, Pac. J. Math., 181(1) (1997),
 159 -- 185.

\bibitem{S2}
B. A. Shipman, The geometry of the full Kostant-Toda lattice of sl(4;C), Journal of Geometry and Physics
33, (2000), 295 -- 325.

\bibitem{K}
B. M. Kostant, On Whittaker vectors and representation theory, Invent. Math. 48 (1978), 101 -- 184.

\bibitem{FT}
L.A. Tahtadzhyan, L.D. Faddeev, Hamiltonian methods in the theory of solitons, "Nauka", (1986).

\bibitem{T}
D.Talalaev, Quantum generic Toda system, arXiv:1012.3296.

\bibitem{F}
W.Fulton, Young Tableaux, Cambridge University Press, 1977.

\bibitem{CSS}
  Y.~B.~Chernyakov, G.~I.~Sharygin and A.~S.~Sorin,
  Bruhat Order in Full Symmetric Toda System,
  arXiv:1212.4803 [nlin.SI].

\bibitem{CS}
Y.~B.~Chernyakov, A.~S.~Sorin,
Explicit Semi-invariants and Integrals of the Full Symmetric $\mathfrak{sl}_n$ Toda Lattice,
  arXiv:1306.1647.

\end{thebibliography}
\end{document}